\PassOptionsToPackage{english}{babel}
\documentclass[aps,pra,
reprint,
superscriptaddress,
 amsmath,amssymb,
floatfix
]{revtex4-2}
\usepackage[dvipsnames]{xcolor}
\usepackage{graphicx}
\usepackage{dcolumn}
\usepackage{bm}
\usepackage{outlines}
\usepackage{float}
\usepackage{textcomp}
\usepackage{hyperref}
\usepackage{subfigure}

\usepackage{tikz}
\usetikzlibrary{angles,quotes}
\usepackage{framed}

\usepackage{titlesec}

\usepackage[normalem]{ulem}

\usepackage{braket}
\newcommand{\derivn}[3][{}]{
    \frac{d^{#1} #2}{d #3^{#1}}
}

\newcommand{\difn}[3]{\frac{d^#3 #1}{d #2^#3}}

\def\edotr{\mbox{\boldmath{$\hat{\bf e}\cdot {\bf r}$}}}

\newcommand{\MetastableState}{2^{3\!}S_1}%
\newcommand{\UpperStateManifold}{3^{3\!}P}%
\newcommand{\LowerStates}{2^{3\!}P_{(0,1,2)}}%
\newcommand{\LowerStateManifold}{2^{3\!}P}%
\newcommand{\SingletState}{2^{1\!}S_0}%
\newcommand{\TO}{\MetastableState- \LowerStateManifold / \UpperStateManifold}%

\usepackage{textcomp}

\begin{document}

\title{
Measurement of a helium tune-out frequency: an independent test of quantum electrodynamics
}

\author{B. M. Henson}
\thanks{These authors contributed equally}
\affiliation{Department of Quantum Science and Technology, Research School of Physics, The Australian National University, Canberra, ACT 2601, Australia}

\author{J. A. Ross}
\thanks{These authors contributed equally}
\affiliation{Department of Quantum Science and Technology, Research School of Physics, The Australian National University, Canberra, ACT 2601, Australia}

\author{K. F. Thomas}
\affiliation{Department of Quantum Science and Technology, Research School of Physics, The Australian National University, Canberra, ACT 2601, Australia}

\author{C. N. Kuhn}
\affiliation{Center for Quantum and Optical Science, Swinburne University of Technology, Melbourne, Australia}

\author{D. K. Shin}
\affiliation{Department of Quantum Science and Technology, Research School of Physics, The Australian National University, Canberra, ACT 2601, Australia}

\author{S. S. Hodgman}
\affiliation{Department of Quantum Science and Technology, Research School of Physics, The Australian National University, Canberra, ACT 2601, Australia}

\author{Yong-Hui Zhang}
\affiliation{State Key Laboratory of Magnetic Resonance and Atomic and Molecular Physics, Innovation Academy for Precision Measurement Science and Technology, Chinese Academy of Sciences, Wuhan 430071, People's Republic of China}

\author{Li-Yan Tang}
\email{lytang@wipm.ac.cn}
\affiliation{State Key Laboratory of Magnetic Resonance and Atomic and Molecular Physics, Innovation Academy for Precision Measurement Science and Technology, Chinese Academy of Sciences, Wuhan 430071, People's Republic of China}

\author{G. W. F. Drake}
\email{gdrake@uwindsor.ca}
\affiliation{Physics Department, University of Windsor, Windsor, Ontario, Canada}

\author{A. T. Bondy}
\affiliation{Physics Department, University of Windsor, Windsor, Ontario, Canada}

\author{A. G. Truscott}
\email{andrew.truscott@anu.edu.au}
\affiliation{Department of Quantum Science and Technology, Research School of Physics, The Australian National University, Canberra, ACT 2601, Australia}

\author{K. G. H. Baldwin}
\email{kenneth.baldwin@anu.edu.au}
\affiliation{Department of Quantum Science and Technology, Research School of Physics, The Australian National University, Canberra, ACT 2601, Australia}

\date{\today}

\begin{abstract}
Despite quantum electrodynamics (QED) being one of the most stringently tested theories underpinning modern physics, recent precision atomic spectroscopy measurements have uncovered several small discrepancies between experiment and theory.
One particularly powerful experimental observable that tests QED independently of traditional energy level measurements is the `tune-out' frequency, where the dynamic polarizability vanishes and the atom does not interact with applied laser light.
In this work, we measure the `tune-out' frequency for the \(\MetastableState\) state of helium between transitions to the $\LowerStateManifold$ and $\UpperStateManifold$ manifolds and compare it to new theoretical QED calculations.
The experimentally determined value of 725\,736\,700\,$(40_{\mathrm{stat}},260_{\mathrm{syst}})$~MHz is within \({\sim} 1.7\sigma\) of theory (725\,736\,252(9)~MHz), and importantly resolves both the QED contributions (\({\sim} 30 \sigma\)) and novel retardation (\({\sim} 2 \sigma\)) corrections.
\end{abstract}

\maketitle

Quantum electrodynamics (QED) describes the interaction between matter and light. It is so ubiquitous that the theory is considered a cornerstone of modern physics.
QED has been remarkably predictive in describing fundamental processes, such as spontaneous emission rates of photons from atoms and the anomalous electron magnetic moment \cite{PhysRevD.91.033006}.
However, as the precision of atomic spectroscopy approaches the part-per-trillion level, discrepancies between such predictions and experiments have come to light, such as the `proton radius puzzle'.
Spectroscopic measurements (of $\mu p$ \cite{Pohl2010}, H \cite{Bezginov1007,Beyer79}, and $\mu d$ \cite{Pohl669}) yield determinations of the proton radius which disagree by up to five standard deviations with other approaches (e+p scattering \cite{ZHAN201159}, and H spectroscopy\cite{PhysRevLett.120.183001}).

Helium is an exemplary testing ground for QED because its simple two-electron structure makes high-precision predictions tractable and testable. Notably helium also presents a nuclear `puzzle', with precision measurement of isotope shifts of the \(\MetastableState \rightarrow \LowerStates \) \cite{PhysRevLett.119.263002} and \(\MetastableState \rightarrow \SingletState \) \cite{Rengelink2018} transitions disagreeing at two standard deviations in the derived nuclear charge radius. Further, recent measurements of the ionization energy for the helium $2^{1\!}S_0$ state \cite{Clausen21} confirm similar discrepancies in the Lamb shift to those recently revealed theoretically \cite{Patkos21}. These `puzzles' raise the possibility that the issue lies with QED itself \cite{refId0}. Thus, we look to challenge QED directly by precision spectroscopy in helium beyond the usual energy interval measurements.

\begin{figure} 
\centering
\includegraphics[width=0.48\textwidth]{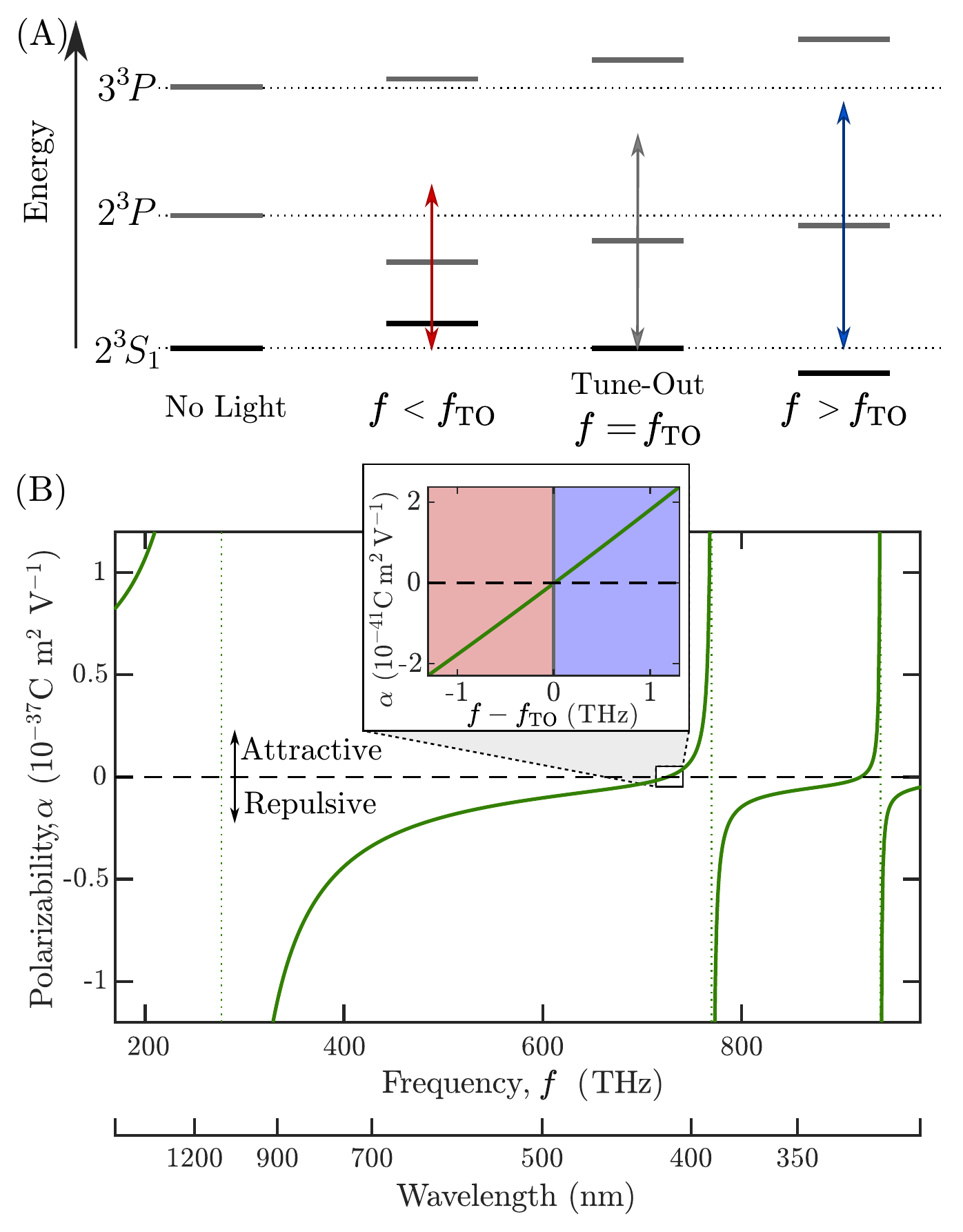}
\caption{\textbf{Tune-out in atomic helium:}
(A) Atomic energy level shift of the dominant state (manifolds) about the tune-out. When an optical field of frequency $f$ (arrows) is applied to the atom the individual
levels shift dependent on the difference between $f$ and the transition frequency. At the tune-out frequency $f_{\mathrm{TO}}$ (middle right), the shifts to the $\MetastableState$ state energy cancel.
Energy spacing and shifts not to scale.
(B) Theoretical frequency dependent polarizability of $\MetastableState$ helium, for a constant light polarization, indicating that the polarizability vanishes near 726~THz, - the tune-out frequency measured in this paper. 
Vertical dotted lines show, from left to right, the transitions to the  $\LowerStateManifold$, $\UpperStateManifold$,$4^{3\!}P$ manifolds. Inset shows the approximately linear polarizability with frequency about the tune-out.
}
\label{fig:schematic} 
\end{figure}

An atom in an optical field experiences an energy shift in proportion to the real part of the frequency dependent polarizability, a fundamental atomic property dictated by the position of energy levels and the strengths of transitions to them (Fig. \ref{fig:schematic}). 
A ‘tune-out’ frequency ($f_\mathrm{TO}$) occurs between transition frequencies at the point where the contributions to the dynamic polarizability ($\alpha(f)$) by all transitions below that frequency are balanced by all those above it ($\alpha(f)=0$) \cite{PhysRevA.75.053612}. 
This balance point is hence fixed by the strength and frequency of every transition in the atomic spectrum and thus provides a precise constraint on the ratio of transition dipole matrix elements (DMEs). 
Similarly, `magic' wavelengths (where the light shift of a transition cancels \cite{Zhang21}, rather than the light shift of a level as is the case for a tune-out wavelength) have yielded absolute and relative determinations of DMEs \cite{PhysRevLett.109.243003,PhysRevA.92.052501}.

As a test of QED, a tune-out frequency is advantageous because it is a null measurement, which does not require calibration of the light intensity or a measurement of excitation probability. These factors have previously limited the precision of direct transition strength measurements \cite{Bouloufa_2009,Vogt2007,PhysRevLett.125.013002}. In comparison, previous tune-out measurements have been successful in measuring QED effects \cite{PhysRevA.92.052501,PhysRevLett.109.243004,PhysRevA.93.022507,PhysRevLett.109.243003,PhysRevLett.115.043004}.

In this work we measure the tune-out of the metastable $\MetastableState$ state of helium (denoted He*) which lies between transitions to the $\LowerStateManifold$ and $\UpperStateManifold$ manifolds (denoted $\TO$) at approximately 726~THz (413~nm). We chose this particular tune-out frequency as the two neighbouring transitions are more than an octave apart in frequency, causing the gradient of atomic polarizability with optical frequency to be very small at the tune-out. Hence, this tune-out frequency is especially sensitive to higher order QED effects. We achieve a 20-fold improvement in the precision over the sole previous measurement \cite{PhysRevLett.115.043004}.

\begin{figure*}[t]
\centering
\includegraphics[width=1\textwidth]{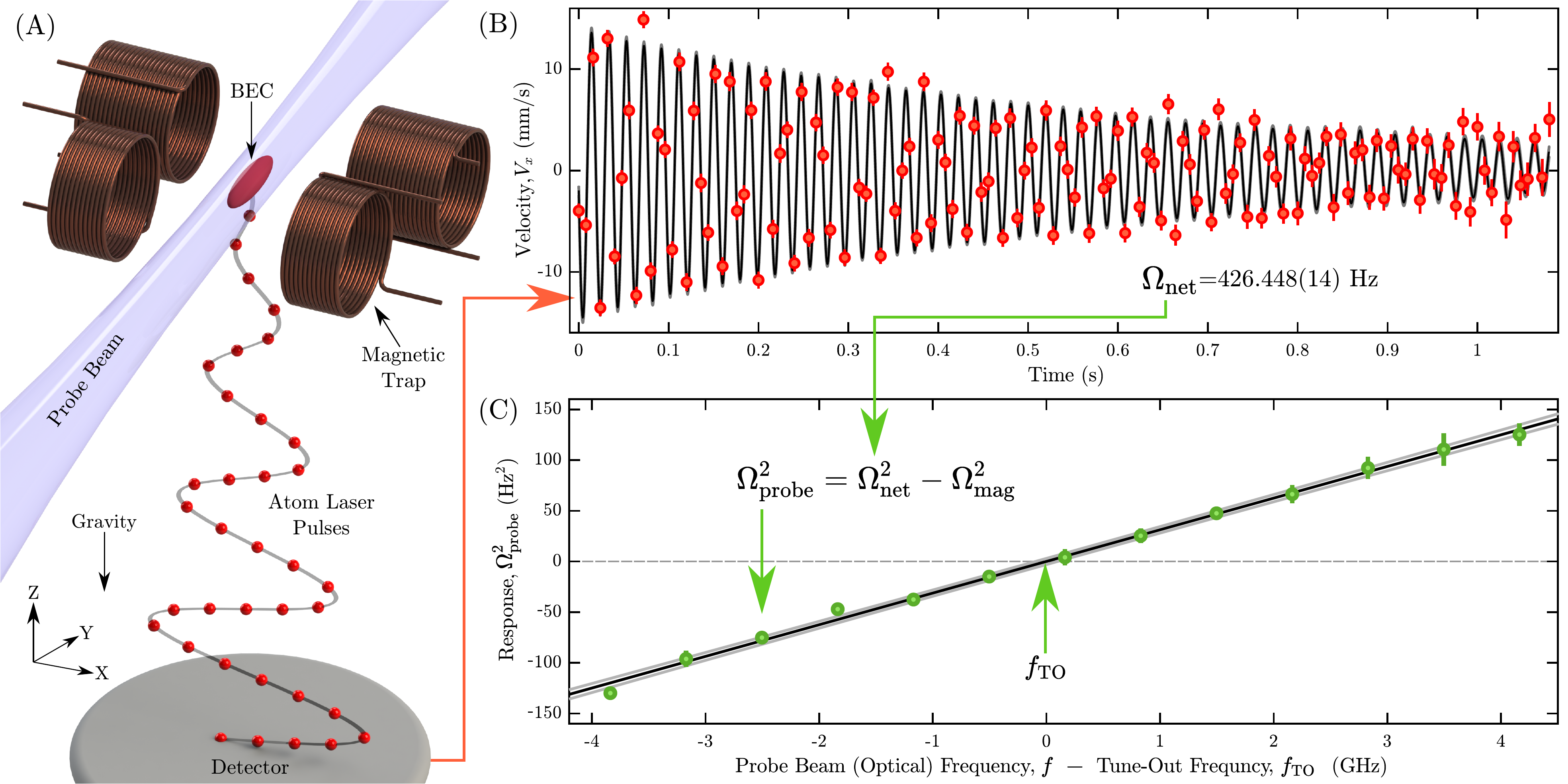}
\caption{
\textbf{Experimental procedure:} Method to determine the tune-out for a fixed probe beam polarization. (A) a magnetically trapped BEC of metastable helium atoms is illuminated with a probe laser beam with an adjustable (optical) frequency. A sequence of atom laser pulses is outcoupled from the BEC to sample the oscillation. (B) The mean velocity of each pulse in the \textit{x}-direction ($v_x$) is used to trace out the oscillation over time (red points) and extract the oscillation frequency with a dampened sine wave fit (solid line). A single experimental realization is shown. (C) The squared probe beam trap frequency (response) is found using a separate measurement of the magnetic trap frequency. This measurement is repeated over a small range of optical frequencies. The tune-out is extracted by finding the \textit{x}-intercept of the response as a function of probe beam frequency using a linear fit (black solid line). Light gray lines show the model \(1\sigma\) confidence intervals. All error bars represent the standard error in the mean.
}
\label{fig:stage_1_processing_schematic} 
\end{figure*}

For an unambiguous comparison we also present a new theoretical estimate of the \(\TO\) tune-out in helium. Following the first prediction \cite{PhysRevA.88.052515} and measurement \cite{PhysRevLett.115.043004} of the tune-out, a vigorous campaign of theoretical studies \cite{PhysRevA.93.052516,manalo2017variational,Drake2019,PhysRevA.99.040502, PhysRevA.99.041803} has reduced the uncertainty in the predicted frequency, which limited comparison with experiment. Our work represents a 10-fold improvement in precision over previous calculations, and whose uncertainty now surpasses the experimental state-of-the-art.

Measuring a tune-out frequency amounts to measuring the potential energy of a light field interacting with an atom, known as an optical dipole potential \cite{grimmOpticalDipoleTraps2000}, and identifying precisely the frequency at which it vanishes (Fig.~\ref{fig:schematic}). The new experimental approach taken here measures the optical dipole potential via changes in the spatial oscillation frequency (also called trap frequency) of Bose-Einsten condensates (BECs) in a harmonic magnetic trap when overlapped with a laser probe beam (Fig.~\ref{fig:stage_1_processing_schematic}). The net potential energy is the sum of a harmonic magnetic potential and a Gaussian optical potential, which is approximately harmonic for the small oscillation amplitudes we consider. In this limit the oscillation frequency is given by $\Omega_{\text{net}}^2=\Omega_{\text{mag}}^2+\Omega_{\text{probe}}^2$ where \(\Omega_{\text{mag}}\), \(\Omega_{\text{probe}}\), and \(\Omega_{\text{net}}\) denote the trap frequency of the magnetic, probe, and combined potentials respectively. For a Gaussian beam profile, as used here, the probe perturbation scales as \(\Omega_{\text{probe}}^2\propto \alpha(f) I\), where I is the intensity of the probe beam. Hence, with the probe beam power stabilized, the difference of squared trapping frequencies \(\Omega_{\text{net}}^2-\Omega_{\text{mag}}^2\propto\alpha(f)\) produces a response which is linearly proportional to the dynamic polarizability. Having measured the transverse and longitudinal profiles of the probe beam, the shift in trapping frequency completely specifies the optical dipole potential.

We determine the trap frequency of our BECs with a novel method which repeatedly samples the momentum of an oscillating BEC with a \textit{pulsed atom laser} \cite{PhysRevLett.78.582,Manning:10} (Fig.~\ref{fig:stage_1_processing_schematic}(A)). Each measurement starts by generating a new He* BEC, which is set in motion by applying a field gradient, and is then depleted over the course of the trap frequency measurement (1.2~s long, Fig.~\ref{fig:stage_1_processing_schematic}(B)). The starting sample of atoms is cooled to ${\sim}80$~nK, well below the critical temperature, to reduce the damping that ultimately limits the interrogation time and, in turn, uncertainty in the trapping frequency. We alternate between measurements of trapping frequency with and without the optical potential to calibrate for any long term drift in \(\Omega_{\text{mag}}\). We then measure the change in (squared) trap frequency due to the probe beam, \(\Omega_{\text{probe}}^2\), as a function of the probe beam (optical) frequency $f$ near the tune-out frequency at  ${\sim} 726$~THz (413~nm). The small laser frequency scan range used in our experiment allows us to determine the tune-out frequency \(f_{\text{TO}}\) via linear interpolation from the measured response of $\Omega_{\text{probe}}^2$ (Fig.~\ref{fig:stage_1_processing_schematic}(C)).

The dynamic atomic polarizability consists of the frequency dependent scalar, vector, and tensor components (\(\alpha^S(f),\alpha^V(f),\alpha^T(f)\) respectively). The total polarizability (and hence the tune-out) also depends on the degree of linear and circular polarization in the atom's reference frame, given by the second and fourth Stokes parameters \(\mathcal{Q_{A}}\) and  \(\mathcal{V}\) respectively, and on the angle $\theta_k$ between the laser propagation direction and the magnetic field vector \cite{LeKien2013}. The tune-out frequency for the \(\MetastableState\) state and arbitrary polarization is: 

\begin{widetext}
    \begin{equation}
    f_{\mathrm{TO}}(\mathcal{Q_{A}}, \mathcal{V}) = f^{S}_{\mathrm{TO}} + \frac{1}{2} \beta^V \cos \left( \theta_k \right) \mathcal{V}  - \frac{1}{2} \beta^T \left[3 \sin^2\left( \theta_k \right) \left(\frac{1}{2} +  \frac{\mathcal{Q_{A}(Q_{L},\theta_{L}})}{2}\right) -1 \right],
    \label{eq:TO_model}
    \end{equation}
\end{widetext}
where \(f^{S}_{\mathrm{TO}}\) is the tune-out frequency for the scalar polarizability $\alpha^S(f)$, \(\mathcal{Q_{A}(Q_{L},\theta_{L}})\) is the second Stokes parameter in terms of the laboratory measurement of the second Stokes parameter \(\mathcal{Q_L}\), and the angle between the lab and atomic frames \(\mathcal{\theta_{L}}\). Here, \( \beta^V\) and  \(\beta^T\) are the vector and tensor polarizabilities divided by the gradient of the scalar polarizability (with respect to frequency) at the tune-out (Sec. \ref{ch:sm.sec:to_comp_lin}).

We measure the tune-out \(f_{\mathrm{TO}}(-1,0)\), corresponding to a linearly polarized light-field whose polarization axis is perpendicular to both the laser propagation and the magnetic field. For this configuration, the sensitivity to \(\theta_{k}\) and \(\theta_\mathcal{L}\) is minimized and the atomic polarizability simplifies to

\begin{equation}
    \alpha(f) = \alpha^S(f) - \frac{1}{2} \alpha^T(f). 
    \label{eq:polarizability_2}
\end{equation}

\begin{figure}
\centering
\includegraphics[width=0.45\textwidth]{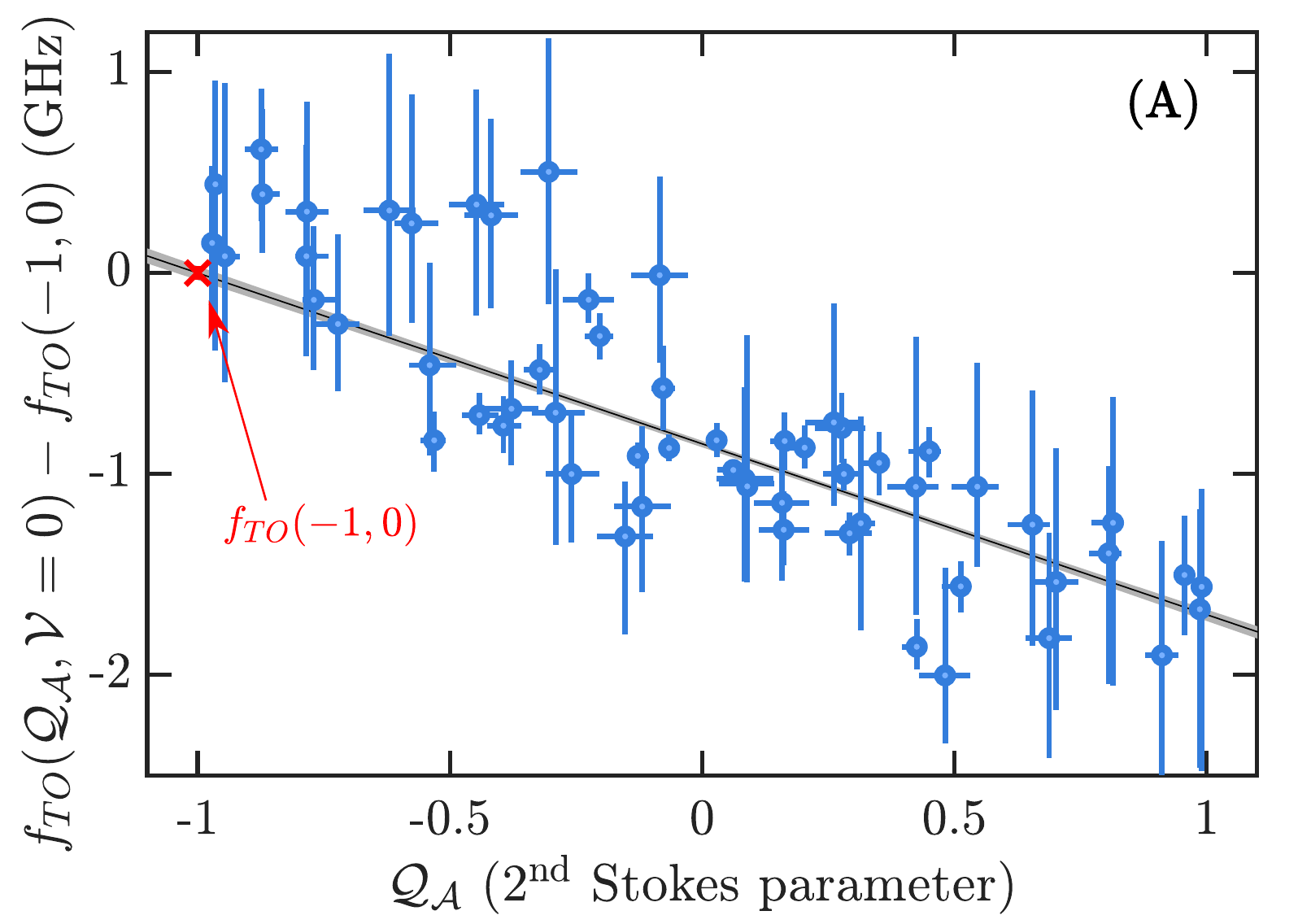}
\includegraphics[width=0.45\textwidth]{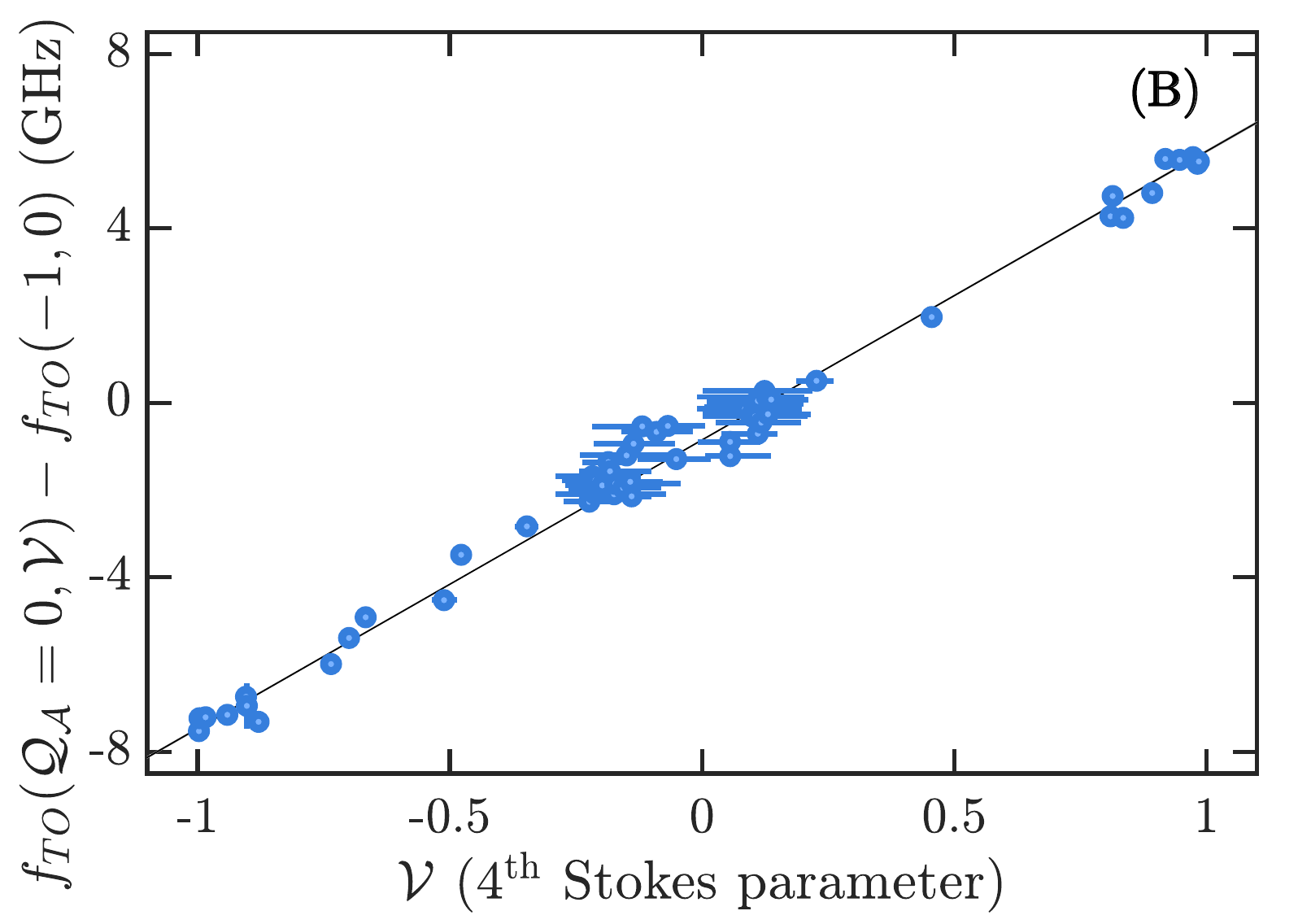}
\caption{\textbf{Tune-out dependence on probe beam polarization:}
(A) Dependence of the measured tune-out on $\mathcal{Q_{A}}$ when interpolated to $\mathcal{V}=0$. 
(B) Dependence of the measured tune-out on $\mathcal{V}$ when interpolated to $\mathcal{Q_{A}}=0$. 
The linear fit to all scans is of the form of Eq.~(\ref{eq:TO_model}), with fit parameters \(f_{TO}(-1,0)=725\,736\,700\,(40)\)~MHz, \(\beta^V \cos(\theta_k)=13\,240\,(70)\)~MHz, \(\beta^T \sin^2(\theta_k)=1\,140\,(20)\)~MHz and $\chi^2$/dof=0.9968.
Horizontal error bars show polarization state uncertainty while vertical error bars show the standard error of the measurement combined with the propagated polarization state uncertainty from the interpolated axis.
For a visualization of the combined dependence see Fig. \ref{fig:full_tune_out}. 
The shaded regions in (A) show the model 1$\sigma$ confidence interval, which is too small to be visible in (B).
The point marked with a red cross shows the reference value $f_\mathrm{TO}(-1,0)$ (error bar not visible at this scale).
} 
\label{fig:pol_TO} 
\end{figure}

We measure \(f_{\mathrm{TO}}(\mathcal{Q_{A}}, \mathcal{V}) \) as a function of the probe beam polarization parameters \(\mathcal{Q_{A}}\) and \(\mathcal{V}\) and interpolate using Eq.~(\ref{eq:TO_model}) to determine \(f_{\mathrm{TO}}(-1,0)\) (Fig.~\ref{fig:pol_TO}). We take the sign of \(\beta^T\) from theory, but use no other predictions in our calculation. Thus, we determine a value of  725\,736\,700\,$(40_{\mathrm{stat}},260_{\mathrm{syst}})$ MHz for the \(f_{\mathrm{TO}}(-1,0)\) tune-out, including systematic effects (Sec. \ref{ch:sm.sec:syst_err}).

The dominant systematic effect in our measurement is the uncertainty in the light polarization. The probe beam passes through a vacuum window before it interacts with the atoms, which may subtly alter the laser polarization relative to measurements made outside the vacuum chamber.
We constrain this error to be less than 200~MHz by measuring the probe beam polarization before entering, and after exiting, the vacuum system (Sec. \ref{ch:sm.sec:syst_err.sub:polz}).

\begin{figure}[t]
    \centering
    \includegraphics[width=0.48\textwidth]{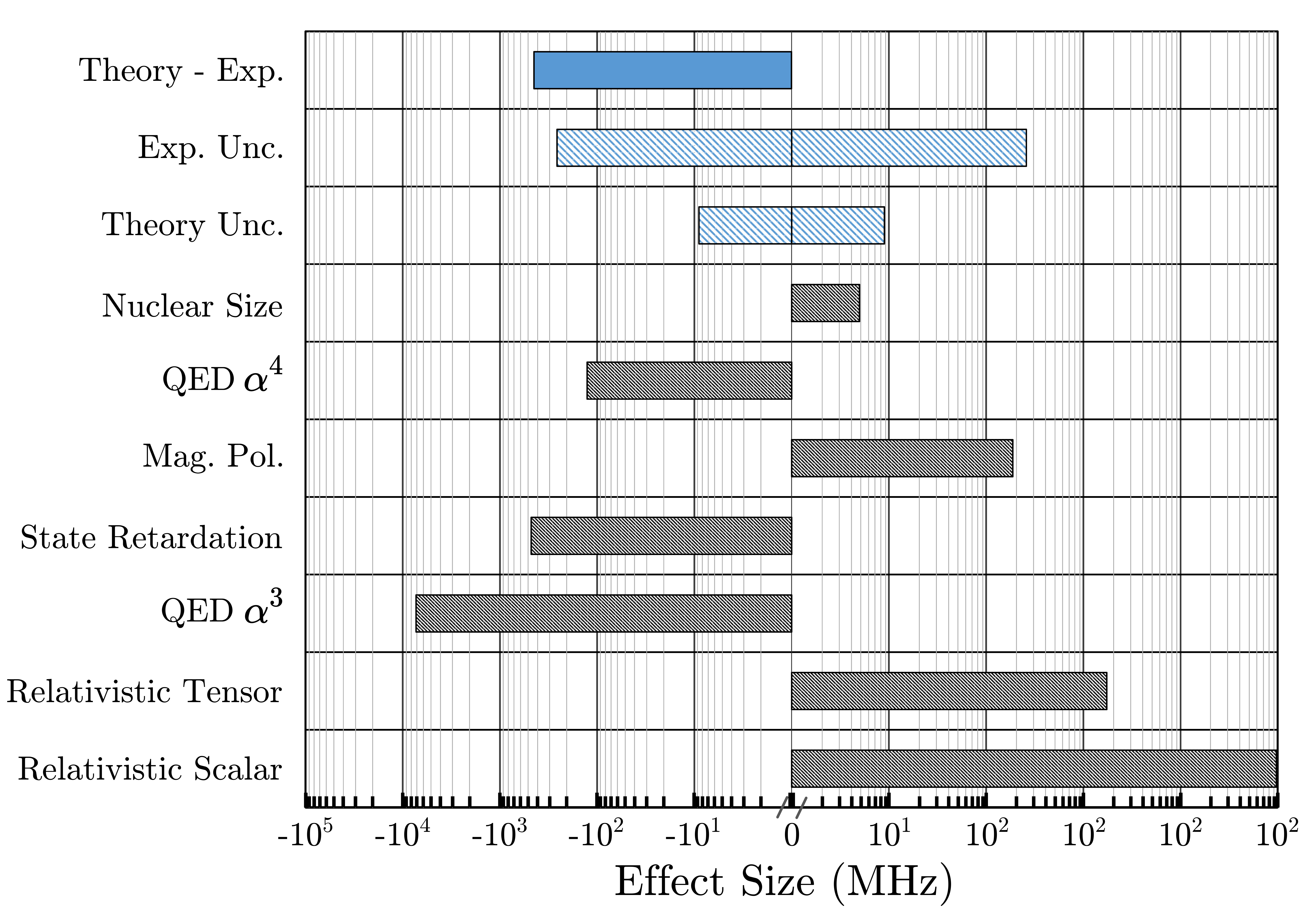}
    \caption{\textbf{Experimental and theoretical sensitivity:} Comparison of uncertainties in the theoretical and experimental determinations of the \(\TO\) tune-out frequency and the various theoretical contributions to the tune-out value.}
    \label{fig:contributions}
\end{figure}

Separately, we improve on the state-of-the-art calculation \cite{PhysRevA.99.040502} of the tune-out frequency by accounting for finite nuclear mass, relativistic, QED, finite nuclear size, and finite wavelength retardation effects \cite{Drake2019, PhysRevA.99.041803} (Sec. \ref{ch:sm.sec:theory}).  
We achieve a 10-fold improvement in precision and find a theoretical value of \(725\,736\,252(9)\)~MHz for \(f_{\mathrm{TO}}(-1,0)\). 
The major contribution to the theoretical uncertainty stems from both the nonradiative QED corrections of order $\alpha^4$ Ry (±6 MHz) and the relativistic scalar part (±6 MHz), for a total of ±9 MHz when added in quadrature, which is an order of magnitude less than the systematic experimental uncertainty (\autoref{ch:sm.tab:exp_results}).
We show a comparison of our experimental and theoretical uncertainties to the main contributions of interest to the theoretical value in Fig.~\ref{fig:contributions}, to demonstrate the contributions to which our measurement is sensitive.

 To summarize, our experimental determination of 725\,736\,700\,$(40_{\mathrm{stat}},260_{\mathrm{syst}})$ MHz is a \(20\)-fold improvement over the first experimental determination, and is 1.7$\sigma$ larger than the theoretical prediction. Our measurement corresponds to a relative precision in oscillator strength ratio of 6 ppm (Sec. \ref{sec:osc_ratio}), 
 which is a factor of two improvement on the previous record \cite{PhysRevA.88.052515}. The combined theoretical and experimental uncertainties (\({\sim} 260\)~MHz) are able to discern the contribution of QED effects (\({\sim} 30 \sigma\)), and are similar to the retardation corrections to the dipole interaction (\({\sim} 2 \sigma\)), but much greater than the contribution of finite nuclear size effects (\(5\)~MHz). Furthermore, our novel method for measuring the dipole potential is able to discern a peak potential energy of as little as 10$^{-35}$J.This is, to our knowledge, the most sensitive measurement of potential energy reported to date (Sec. \ref{sec:pot_sens}).

This is the first measurement to be sensitive to the retardation corrections not normally included in the theory of the frequency-dependent polarizability \cite{Drake2019, PhysRevA.99.041803}. The result is an \({\sim} 1.7 \sigma\) difference between experiment and theory, which takes into account the estimated uncertainty from terms not currently included in the theoretical calculation.
It is notable that by ignoring the retardation correction term – first proposed in Ref.~\cite{PhysRevA.99.041803} and included here in tune-out frequency calculations for the first time - the difference between theory and experiment reduces to \({\sim} 0.1 \sigma\).
If the experimental precision is increased by an order of magnitude, then the effect of the retardation contribution can be more stringently tested.

Future experimental improvements could include more precise laser polarization calibrations, likely using in-vacuum optics, and a finer measurement of the angle between the laser propagation and the magnetic field. This would allow an independent comparison of the predicted and measured scalar, vector, and tensor polarizabilities, providing further information on the structure of the helium atom, and QED theory itself.

Our novel method can be easily applied to other tune-out frequencies in helium, and used as an investigative tool for other problems in QED theory. If the precision of future measurements reach the MHz level, the tune-out frequency could determine the nuclear charge radius of helium. Further improvements and use of our method may thus continue to challenge and elucidate QED theory.


\begin{acknowledgments}

The authors would like to thank Michael Bromley for instructive discussion regarding the hyperpolarizability, Daniel Cocks for careful reading of the manuscript, C. J. Vale and S. Hoinka for the loan of the laser, T.-Y. Shi for helpful discussions regarding the theoretical calculations, 
and Krzysztof Pachucki for helpful correspondence concerning the relativistic and retardation corrections to the tune-out frequency.
This work was supported through Australian Research Council (ARC) Discovery Project Grants No.
DP180101093, 
DP190103021, and 
DP160102337 
as well as Linkage Project No. LE180100142.
K. F. T. and D. K. S. were supported by Australian Government Research Training Program (RTP) scholarships. S. S. H. was supported by ARC Discovery Early Career Researcher Award No. DE150100315. L. Y. T was supported by the National Key Research and Development Program of China under Grant No. 2017YFA0304402 and by the Strategic Priority Research Program of the Chinese Academy of Sciences under Grant No. XDB21030300. G. W. F. D. acknowledges support by the Natural Sciences and Engineering Research Council of Canada and by SHARCNET.
\end{acknowledgments}
\section*{Data Availability}
All primary and processed data along with the associated processing code is available at \cite{tune_out_dataverse,tune_out_zenodo,tune_v2_out_git}.

\bibliographystyle{apsrev4-1}
\bibliography{to_v2}

\clearpage
\onecolumngrid

\renewcommand{\theequation}{S\arabic{equation}}
\renewcommand\thefigure{S\arabic{figure}}
\renewcommand{\thetable}{S\arabic{table}} 
\renewcommand\thesection{S\arabic{section}}

\begin{center}
{\Large \textbf{Supplementary Information}}
\end{center}
\maketitle

\section{Notation}
We propose a new notation $L-U_1/U_2$ (used in this paper) which specifies a tune-out frequency for the occupied state $L$, followed by the two transitions $U_1$,$U_2$ which dominate the polarizability at the specified tune-out frequency. In our case we specify the manifolds which dominate ($\TO$) and neglect the $\MetastableState \rightarrow 3^{3\!}S_1 $ transition as it is sufficiently weak \cite{PhysRevLett.125.013002}. This new notation is similar to the transition notation used in the spectroscopic community.

\section{Method} 	
To measure \(f_{\mathrm{TO}}(-1,0)\) we perform three stages of measurements. Firstly, for a given probe beam polarization and (optical) frequency we make a measurement of the polarizability via the (spatial) oscillation frequency of ultracold helium in the combination of an optical dipole potential of the probe beam and the magnetic trap. 
Secondly, we repeat many of these measurements while varying the probe beam (optical) frequency to find the optical frequency where the polarizability goes to zero (the tune-out frequency, \(f_{TO}(\mathcal{Q_A},\mathcal{V})\) ) for this probe beam polarization. 
Finally, we repeat the second stage at many values of probe beam polarization in order to extract  \(f_{\mathrm{TO}}(-1,0)\), the tune-out for the particular light polarization (in the atomic frame) which we use to compare with theoretical predictions.

\subsection{Trap Frequency Based Polarizability Measurement}
\label{sec:trap_freq_measure}
\subsubsection{Theoretical Basis}
To measure the tune-out we must first be able to be measure the (real part of the) optical polarizability (\(\operatorname{Re}(\alpha)\)) at some given probe beam (optical) frequency. As only the null of this signal is used, absolute calibration is not required. However, the measurement should be linear to allow for linear interpolation of the polarizability as a function of frequency to find the tune-out.  A nonzero polarizability manifests as an optical dipole potential (\(U_{\mathrm{dip}}\)) in proportion to the intensity of the optical field (\(I\)) as \cite{grimm1999optical},
 \begin{align}
    U_{\mathrm{dip}}=-\frac{1}{2 \epsilon_{0} c} \operatorname{Re}(\alpha) I ,
\end{align}
where \(c\) is the speed of light, \(\epsilon_{0}\) is the vacuum permittivity, and \(\alpha\) is the complex polarizability. In this work, we detect the optical dipole potential through a modification to the (spatial) oscillation frequency of ultracold atoms confined in a harmonic magnetic trap when a Gaussian probe beam (oriented along the y axis) is overlapped with the trap minimum. The combined potential is given as
 \begin{align}
    U_{\mathrm{net}}& =-\frac{1}{2 \epsilon_{0} c} \operatorname{Re}(\alpha) 
    \bigg[ 
        \frac{2 P}{\pi w_0^2} \left(\frac{w_0}{w(y)}\right)^2 \exp\left( \frac{-2 (x^2+z^2)}{w(y)^2} \right)
    \bigg]
    + \frac{1}{2} m (x^2 \Omega_{\text{trap},x}^2  +y^2 \Omega_{\text{trap},y}^2  +z^2 \Omega_{\text{trap},z}^2 ), \label{eqn:net_U}
 \end{align}
where \(w(y) = w_0 \sqrt{1+\left( \frac{y}{y_R} \right)^2 }\) is the probe beam waist along the axis of propagation, \(w_0\) is the probe beam waist at the focus, \(y_R\) is the Rayleigh length, \(P\) is the total power of the beam and \( \Omega_{\text{trap},(x,y,z)} \) are the magnetic trap frequencies in the \((x,y,z)\) axes.
To find the net trap frequency we use the expression   
\begin{equation}
\Omega=\frac{1}{2\pi\sqrt{m}} \sqrt{\frac{\partial^{2} U}{\partial x^{2}} \biggr\rvert_{\frac{\partial U}{\partial x}=0}}
\label{eq:omega_from_u}
\end{equation}
for the trap frequency about a local minimum, where \(m\) is the mass of the oscillating particle, and apply it to the net potential in the \(x\)-axis. Applying Eq.~(\ref{eq:omega_from_u}) to Eq.~(\ref{eqn:net_U}) we obtain the net oscillation frequency as
\begin{align}
\Omega_{\text{net}}^2  & = \frac{1}{4\pi^2m} \frac{1}{2\epsilon_{0} c} Re(\alpha) \frac{2 P}{\pi w_0^2} \frac{4}{w_{0}^{2}} 
+  \Omega_{\text{trap},x}^2,
\label{eqn:response_eq}
\end{align}
which can also be expressed as the components from each potential source,
\begin{align}
\Omega_{\text{net}} & =\sqrt{\Omega_{\text{probe}}^{2}+\Omega_{\text{trap},x}^{2}}, \; \; \text{where }\\
\Omega_{\text{probe}}& = \frac{1}{2\pi} \sqrt{\frac{1}{2m \epsilon_{0} c} Re(\alpha) \frac{2 P}{\pi w_0^2} \frac{4}{w_{0}^{2}}} \;.
\end{align}
Here \( \Omega_{\text{probe}} \) represents the spatial oscillation frequency of the probe beam by itself, which becomes imaginary when the trap is repulsive.
Finally we see that a measurement of \( \Omega_{\text{probe}}^2 \) is linearly proportional to the real part of the polarizability. The treatment above is equivalent to a second order Taylor expansion about the trap minimum and provides a good approximation of the behaviour when the oscillation amplitude is small compared to the probe beam waist and when the probe trap frequency is small. For a discussion of the nonlinear effects see section \ref{sec:syst.subsec:lin} below.

\subsubsection{Frequency Measurement}
The trap frequency measurement method in this (and the next) section has been be presented in detail in \cite{henson2022trap}.The pertinent points for this experiment are as follows.
A single measurement of trap frequency begins with the production of a BEC, consisting of $(6\pm1) \times 10^{5}$ metastable \(^4\)He atoms \footnote{variation over the data presented here, which ranges from a minimum of $3 \times 10^{5}$ to a maximum of $8 \times 10^{5}$}, through a combination of laser and evaporative cooling.
The BEC is formed in a biplanar quadrupole Ioffe magnetic trap \cite{Dall2007a} with trapping frequencies given by $(f_{x},f_{y},f_{z})= (426.6(1),55.4(3),428.41(4))$ Hz \footnote{Values in brackets represent variation over many experiments.}.
The probe beam is overlapped with the BEC at the trap minimum (waist radius of  10(2)~\textmu{}m) and aligned along the weak axis ($y$-axis) of the trap, and either turned on in the case of a probe measurement or blocked for a calibration measurement. The probe beam is stabilized in both power and frequency (details below). 
Before measuring the trap frequency a brief (50~\textmu{}s) pulse of current is applied through a small coil near the BEC to induce an oscillation predominantly in the $x$-direction with (initial) amplitude \(\sim30\)~\textmu{}m. 
The momentum of the oscillating BEC is then periodically sampled by weakly out-coupling atoms ($\sim1 \% $) from the magnetic trap every 8~ms with short ($\sim5$~\textmu{}s) pulses of RF radiation resonant with the $m_J=1 \rightarrow m_J=0$ state splitting at the center of the trap ($\sim1.7$ MHz). 
The atoms that are transferred to the $m_J=0$ state are unaffected by magnetic fields, including the trap, and fall 852~mm under gravity to a multi-channel plate and delay line detector \cite{Manning:10} where the large internal energy  ($\sim 20$~eV) of the atoms allows detection with a spatial (temporal) resolution of  $\sim 120$~\textmu{}m ($\sim 3$~\textmu{}s) \cite{PhysRevA.97.063601} and a detection quantum efficiency of ${\sim} 10\%$.
This allows us to reconstruct the initial velocity of each atom in each out-coupling pulse, whose average in each pulse estimates the mean velocity of the BEC at the time of that out-coupling pulse. 
By fitting 130 measurements of the time-dependent velocity with an exponentially damped sine wave we are able to determine the trap frequency to a precision of 10~mHz with a single experimental realization. 
To partially compensate for the decreasing atom number with each pulse, which would otherwise proceed as a geometric series and reduce the signal-to-noise ratio of later pulses, the RF power is increased with pulse number in an approximately exponential manner.

\subsubsection{Reconstructive Aliasing}

The sampling rate of the pulsed atom laser is limited by the momentum width of the BEC in the vertical axis (\(\sim40\)~mm/s, which corresponds to a temporal width at the detector of \(\sim\)4~ms) along with the vertical oscillation amplitude. This presents a challenge for measuring trap frequencies greater than the Nyquist  frequency (62~Hz) of the sampling (at a rate of rate 125~Hz, interval 8~ms) as the Nyquist zone of the original signal is unknown. To find the Nyquist zone of the trap frequency we conduct separate measurements where we vary the sampling rate and measure the change in apparent frequency of the aliased signal. The gradient of the apparent frequency with respect to sampling frequency unambiguously determines the Nyquist zone and in turn the true trap frequency. The high stability of our trap frequency along with the small perturbation of the probe beams allows the net oscillation frequency signal to stay within a single Nyquist zone, even over the entire data acquisition, simplifying the correction. 
In this work, the oscillations in the \(x\)-axis are in the 5\textsuperscript{th} Niquist zone for sampling at 125~Hz and the correction applied to the aliased frequency measured in the oscillation fit is
\begin{equation}
    f_{\text{real}}=3 f_{\text{sampling}} + f_{\text{aliased}}.
\end{equation}
Here \(f_{\text{real}}\) is the true frequency, \(f_{\text{aliased}}\) is the measured aliased frequency and  \(f_{\text{sampling}}\) is the sampling frequency. %

\subsubsection{Probe beam trap frequency}
To extract the squared probe beam trap frequency we use the difference between the squared oscillation frequencies of the combined trap and the the pure magnetic trap (calibration with the probe beam off). Explicitly this reads:
\begin{equation}
    \Omega_{\text{probe}}^{2}=\Omega_{\text{net}}^2 -\Omega_{\text{trap},x}^{2}.  
\end{equation}
Measurements alternate between those with the probe beam on and then off as a calibration.
For a single probe beam measurement the calibration trap frequency (at that time) is taken from an interpolated smoothed (Gaussian kernel, $\sigma=60$~s) model of the calibration measurements in order to correct for trends in the underlying trap frequency and reduce noise. 
This calibration data can be used to provide an estimate of the trap frequency error which combines (true) trap frequency variation along with measurement error. We find a standard deviation of 30~mHz and an overlapping Allan deviation of 18~mHz (at 90~s) in the raw calibration measurements. This values corresponds to a fractional error of 70~ppm and 43~ppm respectively.

\subsection{Tune-Out measurement}
To determine the tune-out frequency (\(f_{TO}\)) for a given polarization state of the probe beam (\( \mathcal{Q_{L}},\mathcal{V} \)), we find the probe beam (optical) frequency $f$ for which the measured probe beam trap frequency \(\Omega_{\text{probe}}^2\) is zero (using the methods described in section \ref{sec:trap_freq_measure} above). 
This is done by measuring \(\Omega_{\text{probe}}^2\) as a function of the probe (optical) frequency in a small range about the tune-out frequency. 
This range is chosen in order to minimize the nonlinearities that are present at large probe beam potentials while still presenting a sufficient signal-to-noise for interpolation of the the linear response. For the data presented here, we used scans out to 4.5~GHz either side of the tune-out.
To perform these scans we change the set point of the probe beam (optical) frequency feedback system every two trap frequency measurements (after each set of one probe-on measurement and one calibration). We step through 13 frequency values over this 9~GHz range. We then use a linear fit to extract the probe beam frequency where \(\Omega_{\text{probe}}^2=0\), which corresponds to the tune-out frequency for this polarization.

\subsubsection{Sensitivity of potential energy measurement}
\label{sec:pot_sens}

A single scan takes approximately 700~s (12~minutes) and consists of 26 trap frequency measurements (BEC productions). We typically achieve an uncertainty in the tune-out of approximately 1.2~$\text{GHz}/\sqrt{N_\text{shots}}$, where \(N_\text{shots}\) refer to the number of BEC's used (either calibration or probe beam measurements). For example, 26 trap frequency measurements (one full scan) gives an uncertainty of $\sim 1.2\;\mathrm{GHz}/\sqrt{26}=235$~MHz. The number of measurements taken to find the tune-out for a given polarization varies from 50 to a few thousand for the data presented here.

Using this method we can infer the peak value of the energy shift imparted by the probe beam onto the atoms. Through Eq.~(\ref{eq:omega_from_u}), a measurement of the probe-induced shift in trap frequency determines the curvature of peak of the Gaussian optical potential energy surface. Along with a measurement of the beam profile at the focus, the inferred curvature completely specifies the geometry of the optical potential. Thus, we can indirectly measure the absolute energy shift in the $\MetastableState$ state with a sensitivity of $1.7\cdot10^{-33}\mathrm{J}/\sqrt{\mathrm{sec}}$, where the time is the probe beam interrogation time. In the case of the probe beam polarization with the lowest frequency uncertainty in $f_\textrm{TO}$, (30~MHz), the minimum potential energy peak we can thus discern with is approximately $10^{-35}\mathrm{J}$ ($U/k_B=3$~pK).

\subsection{Extracting \(f_{\mathrm{TO}}(-1,0)\)}

 The tune-out frequency measurement described in the previous section depends on the light polarization in the atomic frame, which depends in turn on the polarization of the probe beam and the angle between the probe beam propagation and the magnetic polarization axis of the atoms. A comparison of these tune-out measurements to theory would therefore require accurate knowledge of both the probe beam polarization and the magnetic field pointing. In practice we are unable to measure the field pointing with sufficient accuracy to prevent it becoming a limiting uncertainty in our comparison with theory.
 Instead we adopt a different procedure assuming only the stability of the magnetic pointing and use measurements of the tune-out at a variety of polarization states in order to extrapolate to the \(\mathcal{Q_{A}}=-1,\mathcal{V}=0\) polarization state in the atomic frame, giving a tune-out measurement \(f_{\mathrm{TO}}(-1,0)\) that is insensitive to the magnetic field pointing (to first order). 
 
 \subsubsection{Theoretical Basis}
 To understand why \(f_{\mathrm{TO}}(-1,0)\) is independent of field pointing, we start from the expression for the Stark interaction under arbitrary light fields from  \cite[Eq. (19)]{LeKien2013}:
 \begin{equation}
    \alpha(f) = \alpha^S(f) + C \alpha^V(f) \frac{M}{2J} + D \alpha^T(f) \frac{3M^2-J(J+1)}{2J(2J-1)}
    \label{eq:polarizability_full_1}
\end{equation}
where \(\alpha^S(f)\), \(\alpha^V(f)\), and \(\alpha^T(f)\) are the conventional scalar, vector, and tensor polarizabilities respectively. If we assume that the \(B\)-field is pointing along the \textit{z}-axis then the coefficients \(C\) and \(D\) are given by \cite[Eq. (20)]{LeKien2013}:
\begin{align}
    C &= 2 \text{Im}(u_x^* u_y),\\
    D &= 3|u_z|^2 -1.
\end{align}
We can define these constants in terms of experimentally measurable variables,
\begin{align}
     C &= - \mathcal{V_{A}} \cos \left( \theta_k \right), \\
     D &= 3 \sin^2\left( \theta_k \right) \left(\frac{1}{2} +  \frac{\mathcal{Q_{A}}}{2}\right) -1 
\end{align}
where \(\mathcal{V_{A}},\mathcal{Q_{A}}\) are the second and fourth Stokes parameter of the probe beam in the preferred atomic reference frame (see Fig.~\ref{fig:ellipse} for illustration of this particular frame). For the $2^3S_1, M_J=1$ state, which is the focus of this work, we substitute the quantum numbers $J=1$ and $M=1$, leading to the expression,
\begin{equation}
    \alpha(f) = \alpha^S(f) - 
    \frac{1}{2} \alpha^V(f)  \cos \left( \theta_k \right) \mathcal{V_{A}} + 
    \frac{1}{2} \alpha^T(f) \left[3 \sin^2\left( \theta_k \right) \left(\frac{1}{2} +  \frac{\mathcal{Q_{A}}}{2}\right) -1 \right].
    \label{eq:polarizability_full}
\end{equation}
 
 \subsubsection{Tune-Out Component Linearization}
  \label{ch:sm.sec:to_comp_lin}
  
To obtain the dependence of the net tune-out on the Stoke parameters, we perform a Taylor expansion on the dynamic polarizability components \{\(\alpha^S(f)\),\, \(\alpha^V(f)\),\, \(\alpha^T(f)\)\} in terms of frequency about the zero point of \(\alpha^S(f)\), which we denote \(f^{S}_{TO}\),
\begin{align}
    \alpha^J(f) &= \alpha^J(f^{S}_{TO}) + \left. \derivn{\alpha^J}{f}{{}}\right|_{f=f^{S}_{TO}} (f-f^{S}_{TO}) +\frac{1}{2}\left. \difn{\alpha^J}{f}{2}\right|_{f=f^{S}_{TO}}(f-f^{S}_{TO})^2 + ...,\label{eqn:taylor_exp}
\end{align}
where \(J=\{S,V,T\}\). To make the analysis tractable we truncate Eq.~(\ref{eqn:taylor_exp}) to a given order. We wish to truncate the various polarizability components to the lowest order possible, as this makes the final functional form of the tune-out frequency simpler and requires a fewer free parameter fit to the experimental data. This will hence both reduce the fit error and the possibility of having multiple unique local minima in the parameter space.

Theoretically we expect \(\left|\difn{\alpha^{\{V,T\}}}{f}{n} \frac{(\Delta f)^n}{n!}\right|\ll 1\) for \(\Delta f = 16\)~GHz and \(n\geq1\), with the dominant contributions to the net polarizability coming from the first and second derivatives of the scalar polarizability. Hence, we truncate the various components as follows,
\begin{align}
    \alpha^S(f) &\approx \left. \derivn{\alpha^S}{f}{{}}\right|_{f=f^{S}_{TO}} (f-f^{S}_{TO}) +\frac{1}{2}\left. \difn{\alpha^S}{f}{2}\right|_{f=f^{S}_{TO}}(f-f^{S}_{TO})^2 \label{eqn:each_taylor_exp}\\
    \alpha^V(f) &\approx  \alpha^V(f^{S}_{TO}) \\
    \alpha^T(f) &\approx  \alpha^T(f^{S}_{TO}),\label{eqn:each_taylor_exp_vec}
\end{align}
where we have also used \( \alpha^S(f^{S}_{TO})=0\). Substituting these expansions into Eq.~(\ref{eq:polarizability_full}) we obtain,
\begin{align}
    \alpha(f) &\approx \derivn{\alpha^S}{f}{{}}(f-f^{S}_{TO}) + \frac{1}{2} \difn{\alpha^S}{f}{2} (f-f^{S}_{TO})^2
    -\frac{1}{2} \alpha^V(f^{S}_{TO}) \cos \left( \theta_k \right) \mathcal{V}  + 
    \frac{1}{2} \alpha^T(f^{S}_{TO}) \left[3 \sin^2\left( \theta_k \right) \left(\frac{1}{2} +  \frac{\mathcal{Q_{A}}}{2}\right) -1 \right]  . \label{eq:main_to}
\end{align}
We wish to determine the quantity \(f_{TO}\), at which the net polarizability vanishes, \(\alpha(f_{TO})=0\). As changes in the total polarization \(\alpha(f)\) near the tune-out come predominantly from the scalar polarizability we have \(f_{TO}\approx f^{S}_{TO}\). Thus, we can assume that our truncate Taylor expansions of the polarizability terms [Eqs.~(\ref{eqn:each_taylor_exp}-\ref{eqn:each_taylor_exp_vec})], and hence Eq.~(\ref{eq:main_to}), are valid over the range of interest. Furthermore, we note that we try a fit to the data both including and excluding the quadratic term in Eq.~(\ref{eqn:each_taylor_exp}) and find that both models reproduce the same fit values within uncertainties. We thus determine that for simplicity we can simplify Eq.~(\ref{eqn:each_taylor_exp}) to linear order. Setting \(f=f_{TO}\) and solving Eq.~(\ref{eq:main_to}) we find our tune-out equation, 
 \begin{align}
    f_{TO} &= 
    f^{S}_{TO}
    +\frac{1}{2}\beta^V  \cos \left( \theta_k \right) \mathcal{V_{A}}
    - \frac{1}{2}\beta^T  \left[3 \sin^2\left( \theta_k \right) \left(\frac{1}{2} +  \frac{\mathcal{Q_{A}}}{2}\right) -1 \right], \label{eqn:tune_out_eq} 
\end{align}
where
 \begin{align}
  \beta^V &= \alpha^V(f^{S}_{TO}) \bigg/ \left. \derivn{\alpha^S}{f} \right|_{f=f^{S}_{TO}} \mathrm{, and}\\
  \beta^T &= \alpha^T(f^{S}_{TO}) \bigg/ \left. \derivn{\alpha^S}{f} \right|_{f=f^{S}_{TO}} .
 \end{align}
Note that the choice to expand about \(f^{S}_{TO}\) in Eq.~(\ref{eqn:taylor_exp}) is somewhat arbitrary, any frequency sufficiently close to the net tune-out \(f_{TO}\) can be chosen and will produce an equivalent functional form to Eq.~(\ref{eqn:tune_out_eq}).  We chose \(f^{S}_{TO}\) as it simplifies the interpretation of the final equation. If we set \(\mathcal{V_{A}}=0\) and \(\mathcal{Q_{A}}=-1\) we obtain \(f_{TO}(-1,0) = f^{S}_{TO} + \frac{1}{2} \beta^T\) which is the tune-out frequency for the dynamic polarizability \(\alpha(f) = \alpha^S(f) - \frac{1}{2}  \alpha^T(f)\), and is hence independent of the magnetic field pointing. We find further support for this analysis from the experimental data, as it provides a good fit to Eq.~(\ref{eqn:tune_out_eq}) (see Fig.~\ref{fig:full_tune_out} and Fig.~3 in the main text).

 \subsubsection{Polarization in the Atomic Reference Frame}
 We measure the probe beam polarization parameters \(\mathcal{V_{L}},\mathcal{Q_{L}}\) in the laboratory basis using a high extinction rotating polarizer \footnote{Glan-Thompson, extinction ratio $>10^{5}:1$} and the power ratio technique. The polarization parameters are given by
 \begin{align}
     \mathcal{Q_{L}} &=\frac{p_{\mathrm{max}}-p_{\mathrm{min}}}{p_{\mathrm{max}}+p_{\mathrm{min}}} \cos(2\theta_{\mathrm{min}}), \\
     |\mathcal{V}_{\mathcal{L}}| &= \frac{2\sqrt{p_{\mathrm{min}}p_{\mathrm{max}}}}{p_{\mathrm{min}}+p_{\mathrm{max}}},
\end{align}
where \(p_{\mathrm{max}}\) (\(+p_{\mathrm{min}}\)) is the maximum (minimum) power transmitted and $\theta_{\mathrm{min}}$ is the polarizer angle of minimum power transmission. The sign of \(\mathcal{V}_{\mathcal{L}}\), corresponding to the handedness of the circular component, is determined using a rotating quarter wave plate technique. The second polarization parameter is invariant under transformation into the atomic reference frame, hence \(\mathcal{V_{L}}=\mathcal{V_{A}}\) is hereafter denoted \(\mathcal{V}\). The fourth polarization parameter is transformed into the atomic reference frame by a rotation by \(\mathcal{\theta_{L}}\) (see Fig.~\ref{fig:ellipse}) around the probe beam axis,
\begin{equation}
 \mathcal{Q_{A}} =\frac{p_{\mathrm{max}}-p_{\mathrm{min}}}{p_{\mathrm{max}}+p_{\mathrm{min}}} \cos(2(\mathcal{\theta_{L}}+\theta_{\mathrm{min}})) ,
\end{equation}
which corresponds to a rotation about the probe beam to align the laboratory polarizer angle origin with the component of the magnetic field vector pointing radially from the probe beam. In practice \(\mathcal{\theta_{L}}\) cannot be directly measured with sufficient accuracy and is used as a free fit parameter, as described in the next section. 

These equations are sufficiently nonlinear that we use monte-carlo error propagation 
to find the (asymmetric) 68\% confidence intervals of the polarization parameters in the atomic frame. The confidence intervals in $\mathcal{V}$, $\mathcal{Q_{A}}$ are found from the percentiles of many numerical experiments where the measured analyzer values are perturbed by a normal random variable with a standard deviation equal to the estimated measurement error in that parameter.

 \tikzset{
   pics/.cd,
   vector out/.style={
      code={
         \draw[#1] (0,0)  circle (1) (45:1) -- (225:1) (135:1) -- (315:1);
      }
   }
}
\tikzset{
   pics/.cd,
   vector in/.style={
      code={
        \draw[#1] (0,0)  circle (0.25);
        \fill[#1] (0,0)  circle (.05);
      }
   }
}
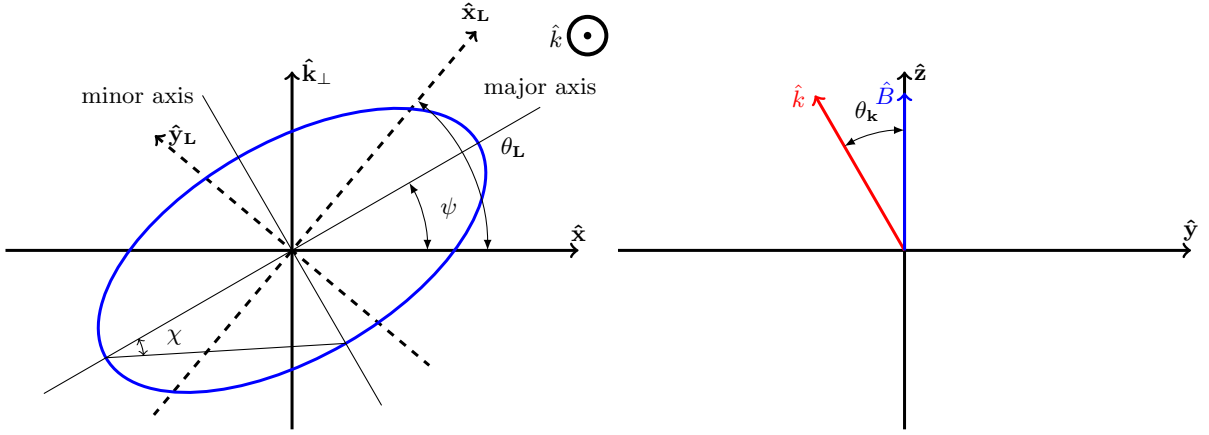
\begin{figure}
\centering
\begin{tikzpicture}[declare function={c=1.4;a=4/c;b=2/c;alpha=30;}]
\begin{scope} 
 \draw[<-,line width=0.4mm] (0,5*b/3) node[right]{$\mathbf{\hat{k}_\perp}$} -- (0,-5*b/3);
 \draw[<-,line width=0.4mm] (4*a/3,0) node[above]{$\mathbf{\hat{x}}$} -- (-4*a/3,0);
\end{scope}

\begin{scope}[dashed,rotate=50] 
 \draw[<-,line width=0.4mm] (0,5*b/3) node[right]{$\, \mathbf{\hat{y}_L}$} -- (0,-5*b/3);
 \draw[<-,line width=0.4mm] (4*a/3,0) node[above]{$\mathbf{\hat{x}_L}$} -- (-3*a/3,0);
\end{scope}

\begin{scope}[rotate=alpha] 
 \draw[line width=0.4mm,blue] (0,0) circle (a and b); 
 \draw (0,5*b/3) node[left]{minor axis} --  (0,-5*b/3) ;
 \draw (4*a/3,0) node[above]{major axis} -- (-4*a/3,0);
  \draw (-a,0) -- (0,-b); 
    \coordinate (a) at (0,0);
    \coordinate (b) at (-a,0);
    \coordinate (c) at (0,-b);
 \pic ["$\mathbf{\chi}$",draw, <->,radius=0.8cm, angle eccentricity=1.9]{angle = c--b--a};
 \path 
 ({a*cos(atan(-(b/a)*tan(alpha)))},{b*sin(atan(-(b/a)*tan(alpha)))}) coordinate (aux1)
 ({a*cos(atan((b/a)*cot(alpha)))},{b*sin(atan((b/a)*cot(alpha)))}) coordinate (aux2) 
 ({-a*cos(atan(-(b/a)*tan(alpha)))},{-b*sin(atan(-(b/a)*tan(alpha)))}) coordinate (aux3)
 ({-a*cos(atan((b/a)*cot(alpha)))},{-b*sin(atan((b/a)*cot(alpha)))}) coordinate (aux4); 
\end{scope}
\path (0:1) coordinate (A) (0,0) coordinate[label={[xshift=0.3em]below left:{$ $}}] (O)
 (alpha:1) coordinate (C) 
pic ["$\mathbf{\psi}$",draw,latex-latex,angle radius=1.8cm,angle eccentricity=1.2] {angle = A--O--C};
\path (0:1) coordinate (B) (0,0) coordinate (D)
 (50:1) coordinate (E) 
pic ["$\mathbf{\theta_L}$",draw,latex-latex,angle radius=2.6cm,angle eccentricity=1.25] {angle = B--D--E};
\path (5.5/c,4/c) pic {vector in={line
width=1.5pt}} node[left]{$\hat{k}\, \, \, \,$};
\end{tikzpicture}
\begin{tikzpicture}[declare function={c=1.4;a=4/c;b=2/c;alpha=30;}]
\begin{scope} 
 \draw[<-,line width=0.4mm] (0,5*b/3) node[right]{$\mathbf{\hat{z}}$} -- (0,-5*b/3);
 \draw[<-,line width=0.4mm] (4*a/3,0) node[above]{$\mathbf{\hat{y}}$} -- (-4*a/3,0);
\end{scope}
\begin{scope}[rotate=alpha] 
 \draw[<-,line width=0.4mm,red] (0,5*b/3) node[left]{$\hat{k}$} --  (0,0);
 \draw[<-,line width=0.4mm,blue] (0.5*2.1,2.1*0.866) node[left]{$\hat{B}$} -- (0,0);
\end{scope}
\path (90:1) coordinate (A) (0,0) coordinate[label={[xshift=0.3em]below left:{$ $}}] (O)
 (120:1) coordinate (C) 
pic ["$\mathbf{\theta_k}$",draw,latex-latex,angle radius=1.6cm,angle eccentricity=1.2] {angle = A--O--C};
\end{tikzpicture}
\caption{Diagram showing the various parameters used in Eq.~(\ref{eq:main_to}). (left) Shows the Stokes ellipse (blue) where \(\mathcal{Q} = \cos(2\psi) \cos(2\chi)\) and \(\mathcal{V} = \sin(2\chi)\), and \((\hat{x}_L,\hat{y}_L)\) represents the lab reference frame. Note that \(\hat{k}_\perp=\hat{y}\cos(\theta_k)+\hat{z}\sin(\theta_k)\) and in this case the probe beam wavevector \(\hat{k}\) is out of the page. (right) Shows how the magnetic quantization axis is assumed to be along the \(z\)-axis, with the plane spanned by it and the probe beam wavevector forming the \(zy\)-plane.}
\label{fig:ellipse}
\end{figure}
 \subsubsection{Fitting}
 
To find \(f_{\mathrm{TO}}(-1,0)\) we measure the tune-out frequency and \(\mathcal{Q_{L}},\mathcal{V}\) over a range of $\lambda/2$, and $\lambda/4$ wave-plate angles (75 combinations used here) and then fit the above model [Eq.~(\ref{eqn:tune_out_eq})] to this set of \(\{\mathcal{Q_{L}},\, \mathcal{V},\, f_{TO}\}\) data using the free parameters $f^{S}_{TO}$,\, $\theta_{\mathcal{L}}$,\, $\theta_{k}$,\, $\beta^V$, and $\beta^T$ (see Fig.~\ref{fig:full_tune_out} for fit of full model with polarization data from pre and post vacuum chamber). This free fit is unable to fully constrain the free fit parameters, critically, giving equal agreement between either polarity of \( \beta^T \) and thus preventing a determination of \(f_{\mathrm{TO}}(-1,0)\). 
To eliminate one of these cases, we introduce a constraint on the sign of  \( \beta^T \) using  measurements and simulations of the magnetic field pointing and theoretical atomic structure calculations, both of which agree with the sign constraint \( \beta^T >0 \).
With this constraint added, we use an uncertainty-weighted fit to find \(f_{\mathrm{TO}}(-1,0)\) by evaluating the resulting model at \(\mathcal{Q_{A}}=-1,\mathcal{V}=0\). The statistical error in the calculated \(f_{\mathrm{TO}}(-1,0)\) is determined with a bootstrapping technique wherein the constrained fitting procedure is repeated on subsets of the data to estimate the uncertainty in the full fit \cite{bryce_bootstrap_error_code}. 
Four of the fit terms ($\theta_{\mathcal{L}}$, $\theta_{k}$, $\beta^V$, and $\beta^T$) are interdependent (reflected in their nondiagonal covariance matrix) and therefore the result of this fit cannot be used to find these values without extra information (such as measuring all but one such parameter). However, this does not effect the prediction of \(f_{\mathrm{TO}}(-1,0)\).

For display in Fig.~\ref{fig:full_tune_out} the $\mathcal{Q_{A}}$ value is calculated using the fit $\theta_{\mathcal{L}}$ and $\mathcal{Q_{L}}$ and the measured \(f_{TO}\) is displayed as a function of $\mathcal{Q_{A}}$, $\mathcal{V}$. In Fig.~\ref{fig:full_tune_out} we also display the tune-out calculated for polarization data sets taken before and after the vacuum chamber.

\subsubsection{Simplified Explanation}
It can be helpful to consider this process for a simplified system with only linear polarization. In this case the measured tune-out will depend sinusoidally on the angle of the input polarization (\(\theta_{\mathcal{L}}\)). The above method is equivalent to using a sinusoidal fit in order to extract the maximum tune-out value in this dependence, corresponding to the \(f_{\mathrm{TO}}(-1,0)\). The choice of taking the maximum is equivalent to constraining \( \beta^T \). The above method is a natural extension of this simplified approach to also account for the circular component of the light field.


\section{Experimental Details}
The general experimental setup is depicted diagrammatically in Fig.~\ref{fig:exp_diagram}. Each section of the experimental setup is described in detail below.
\subsection{Technical Details}
\subsubsection{Laser System}
The laser system produces the monochromatic optical radiation which forms the probe beam. The core components of the system are a tunable titanium sapphire laser (m-Squared SolsTiS-PSX), external doubling cavity (m-Squared ECD-X) and high precision wavemeter (High Finesse WS8-2). The system delivers up to 150~mW of tightly focused (10~\textmu{}m radius), frequency-tunable optical radiation, stabilized in both power and frequency, into the experimental chamber.

\begin{figure}
    \centering
    \includegraphics[width=0.9\textwidth]{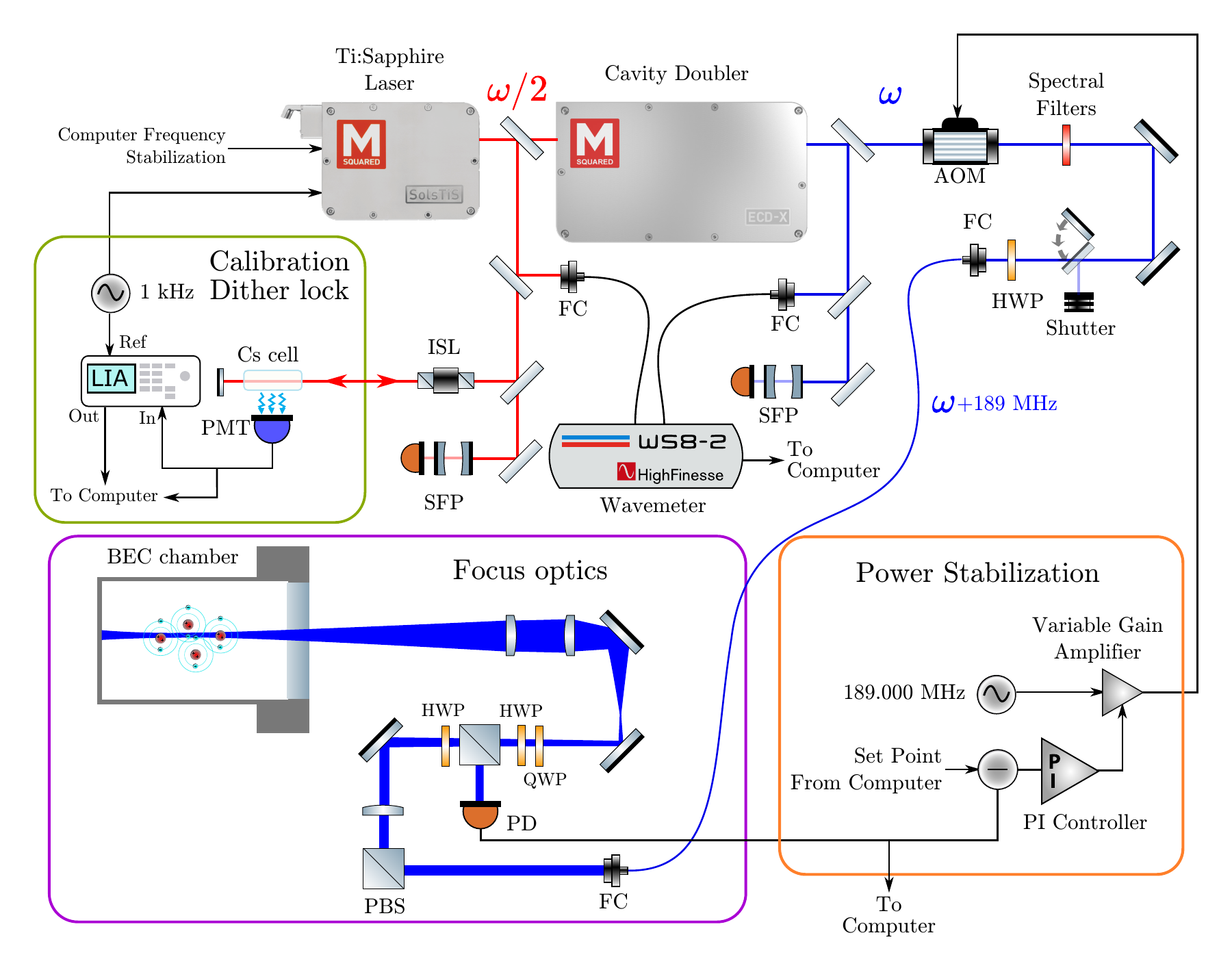}
    \caption{Schematic of the optics system. The AOM diffracts into the +1 order with other orders blocked. Abbreviations:  SFP - scanning Fabry-Perot cavity, FC- fiber coupler, ISL-optical isolator, LIA-lock-in amplifier, PBS- polarizing beam splitter, PD- photodiode, AOM- acousto-optic modulator. }
    \label{fig:exp_diagram}
\end{figure}

\subsubsection{Power Feedback}
The power after the delivery fiber is measured with an amplified photodiode which is used to control the drive power to an acousto-optic modulator (AOM). The feedback loop has a 3dB bandwidth of 170 kHz and stabilizes the beam power to an accuracy of $ 3\times10^{-3}$ relative to the mean power. 

\subsubsection{Wavemeter Feedback}
During normal operation the laser frequency is continuously stabilized using a software-based control loop. The system reads the optical frequency from the wavemeter, passes this through a software based PI controller, and issues the resulting feedback commands to the laser at a rate of $\sim 14$~Hz. The system has sufficient bandwidth to achieve a $\sim900$~ms rise time (10\% to 90\%) for a $\sim300$~MHz frequency step, and during interrogation we measure a typical (in-loop) standard deviation of 170~kHz.
The set point in this control loop is stepped every two BEC cycles (one interrogation and one calibration) to scan out a measurement of the tune-out frequency. The stabilization of the laser frequency uses measurements of the red side of the laser system (predoubling cavity) such that interruptions to the doubling cavity do not impact the optical frequency feedback.


\subsubsection{Wavemeter Calibration}
To provide an absolute calibration of the wavemeter we use the Doppler-free two-photon $6^{2}S_{1/2} (F=3) \rightarrow 8^{2}S_{1/2} (F=3)$ and $6^{2}S_{1/2} (F=4) \rightarrow 8^{2}S_{1/2} (F=4)$ transitions in cesium around 364.5~THz (822.5~nm). We split off a small fraction ($\sim$50~mW) of the red light generated by the Ti:Sapphire laser (before the doubling cavity), pass it through a warm (50$^{\circ}$~C) cesium cell with a beam waist of $\sim0.5$~mm, and then reflect it backwards along its path.
We detect the excitation of the transition using the blue florescence from the radiative cascade \cite{HAGEL19991}, with a blue-sensitive (red-blind) photomultiplier tube (PMT). Previous measurements \cite{Wu:13,Fendel:07} have precisely measured the $F=3\rightarrow3$ and $4\rightarrow4$ transitions to be 364\,507\,238.363(10) and 364\,503\,080.297(10) MHZ respectively, and have demonstrated an insensitivity to environmental conditions, which makes these transitions suitable as a secondary frequency standard. 

To calibrate the wavemeter we disable the usual software based wavemeter feedback to the laser and instead stabilize the laser using one of these transitions. To produce a derivative error signal suitable for feedback we modulate the frequency of the Ti:Sapphire laser (frequency deviation $<50$~kHz, modulation frequency $\sim1$~kHz) and detect the resulting modulation in the PMT current with a lock-in amplifier. This analog error signal is continuously read by a software based PID controller which sends adjustment commands to the laser controller (rate $\sim20$~Hz) to maintain the the laser frequency at the maximum of the fluorescence.

\begin{figure}
    \centering
    \includegraphics[width=0.6\textwidth]{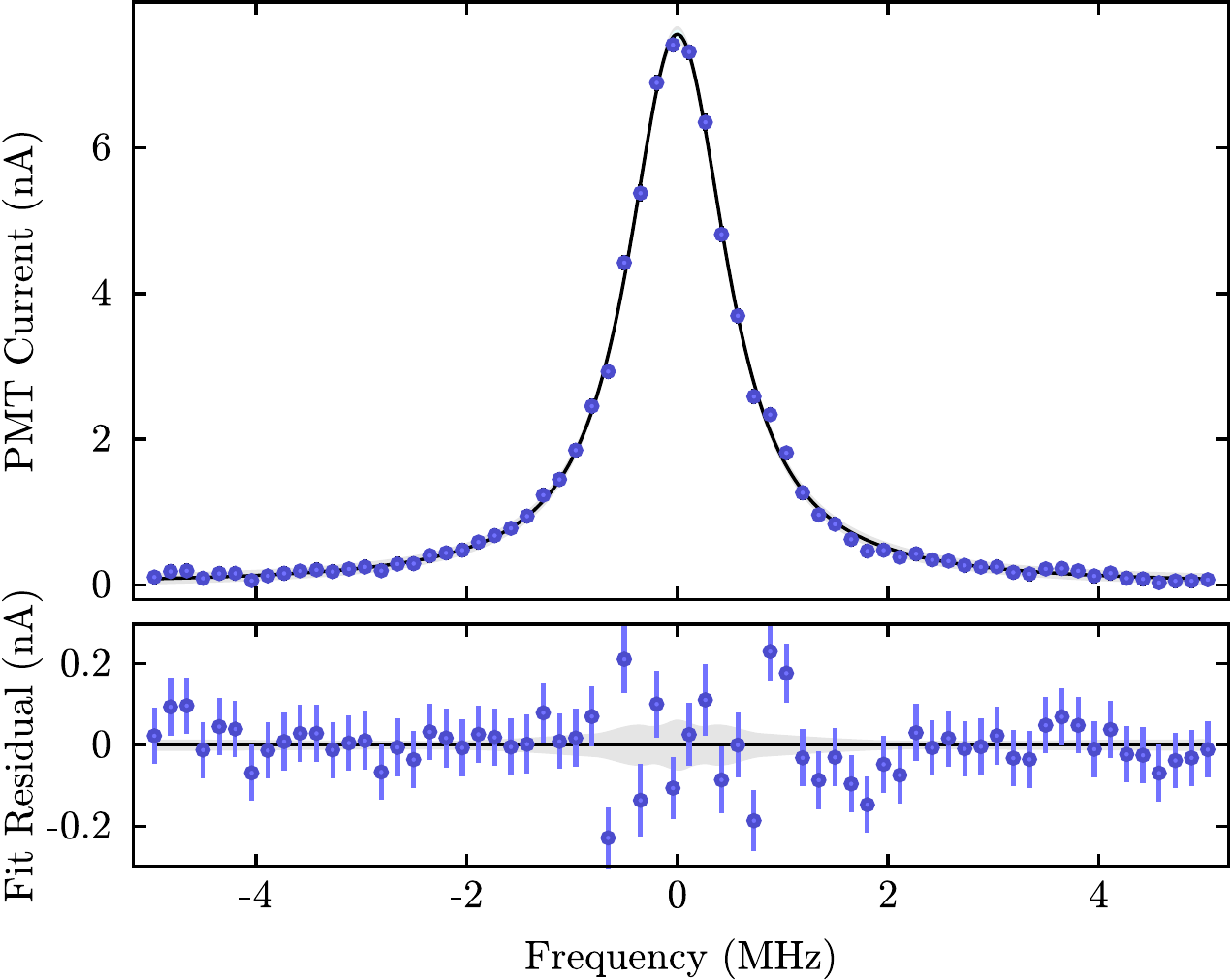}
    \caption{(Top) single scan of PMT current vs. optical frequency relative to the two photon transition $6^{2}S_{1/2} (F=3) \rightarrow 8^{2}S_{1/2} (F=3)$. A Voigt fit is shown as the black line for comparison, with fit parameters are $\sigma=0.18(3)$~MHz $\gamma=0.49(3)$~MHz. This scan took a total of \(75\)~s. (Bottom) Residuals of the fit model, shaded region shows the model 1$\sigma$ confidence intervals.
    }
    \label{fig:2p_scan_single}
\end{figure}

As a verification of the calibration procedure we then re-engage the wavemeter based laser feedback system and measure the PMT current as a function of the frequency set point. We fit these data with a Lorentzian profile to extract the transition frequency and verify the calibration procedure (see Fig.~\ref{fig:2p_scan_single}). After calibration we find that measurements of both $F=3\rightarrow3$ and $F=4\rightarrow4$ transitions give frequencies within 50~kHz of the reference values.
Calibrations are carried out every few days as the ($\sim100$~mK) temperature stability of our laboratory reduces the thermal drift of the wavemeter.
Based on previous systematic studies of these transitions \cite{Wu:13,Fendel:07} and the conditions used for calibration, we believe that the systematic error of this calibration procedure ($<100$~kHz) is well below the absolute accuracy of the wavemeter over the measurement range used in the this work (2~MHz within 3~THz (2~nm) of calibration \cite{wstechnical}). As the wavemeter measurement and calibration is carried out on the red side of the laser system before the doubling cavity the absolute accuracy of the frequency of the delivered (blue) light is doubled to 4~MHz, well below other systematic uncertainties.


\subsection{Data Vetos}
To protect the integrity of the final data, we implemented a number of `veto' protocols in the data analysis which discard polarizability measurements when certain desirable experimental conditions are not met. While stringent, these conditions remove only a negligible amount of data (less than one in $10^{4}$ of all shots in the entire dataset). 

\subsubsection{Laser Single Mode}
We have found that it is possible (although rare) for the titanium sapphire laser to spontaneously run multimode (with multiple optical frequencies in the output) which prevents accurate optical frequency measurement by the wavemeter and in turn accurate polarizability measurement. To eliminate these data we monitor the (folded) optical spectrum using a scanning Fabry-Perot cavity (Thorlabs SA200-5B). The cavity length is scanned across two full free-spectral ranges of the cavity, with a sawtooth voltage applied to the piezoelectric actuator which controls the cavity length at frequency of 20 Hz. 
Both the cavity piezoelectric actuator and photodiode voltage are acquired around (during and $\sim0.2$~s before/after) the probe interrogation to verify that the laser is running single mode. 
We detect a multimode condition based on the separation of peaks in the photodiode voltage. During single mode operation peaks will be separated by the cavity free spectral range with only detector noise in-between. A multimode laser appears as peaks spaced closer than this nominal separation, activating the veto condition. We set a conservative threshold for peak detection which corresponds to $1.5\times10^{-3}$ of the peak transmission intensity noting that all multimode conditions we have observed have intensities orders of magnitude greater than this level.
\subsubsection{Probe Power}
We employ a check on the probe beam power as measured with the probe beam photodiode which measures the power of the beam just before it enters the experimental chamber, as shown in Fig.~\ref{fig:exp_diagram}. This ensures that the power feedback system is operating at set point in all measurements used in the analysis. Restrictive thresholds are set on the mean ($<0.02$ fractional difference to set point), standard deviation ($<0.05$ of set point) and maximum difference to set point ($<0.03$ of set point) of the power during interrogation.
\subsubsection{Optical Frequency}
For our analysis the average value of the optical frequency as measured by the wavemeter (used for feedback) is used for each interrogation. We apply thresholds on the standard deviation ($<3$~MHz) and range ($<5$~MHz) of the measured optical frequency during interrogation.
\subsubsection{Atom number}
To prevent data with erroneously low atom numbers from corrupting measurements, we apply a requirement that atom number is at least $2\times10^{5}$. The systematic shift due to atom number is very small, which allows data taken across varying atom numbers to contribute to the overall result.

\section{Systematic Shifts}
 \label{ch:sm.sec:syst_err}
The systematic shifts to our value for the tune-out are summarised in \autoref{ch:sm.tab:exp_results}, and detailed individually below.
\begin{table}[t]
\centering
\begin{ruledtabular}
\begin{tabular}{l|r|r}
Term              & Estimate &  Uncertainty \\
\hline
Measured Value      & 725\,736\,810             & 40      \\
Polarization        & & \\
\, \, - Birefringence & -100                   & 200   \\
\, \, - Beam Anisotropy & 0                   & 150   \\
Method Linearity    & 24                   & 30       \\
Hyperpolarizability           & -30                   & 50   \\
Broadband Light     & 0                     & 30      \\
DC Electric field   & 0                     & \(\ll 1\) \\
Wave-meter          & 0                     & 4    \\
Mean-Field          & 0                     & \(\ll 1\) \\
\hline
Total               &  725\,736\,700            &  260
\end{tabular}
\end{ruledtabular}
\caption{Contributions to measured tune-out frequency with associated systematic uncertainties (MHz). The measured value is found using only polarization data measured after the vacuum chamber. The polarization row gives the average of the tune-out frequencies calculated using polarization data pre and post vacuum chamber relative to the measured value, where the uncertainty is the discrepancy between these values. Note that uncertainties are added in quadrature.}
\label{ch:sm.tab:exp_results}
\end{table}

\subsection{Polarization}
 \label{ch:sm.sec:syst_err.sub:polz}
 The method we have detailed above to extract \(f_{\mathrm{TO}}(-1,0)\) relies on the accurate measurement of the probe beam polarization when it interacts with the atoms. However, as we do not have polarization optics inside our vacuum system; we must infer it from measurements outside the vacuum system. In this section we make estimates of two effects: first, the unknown birefringence of the vacuum windows, and second, the variation in polarization across the beam.
 
 \subsubsection{Birefringence}
 To estimate the first effect we measure the probe beam polarization before it enters the vacuum system and again after it exits through a second window. The estimated error in the value of \(f_{\mathrm{TO}}(-1,0)\) from the vacuum entry window is constrained by conducting the \(f_{\mathrm{TO}}(-1,0)\) fitting procedure with both measured polarization sets (see Fig.~\ref{fig:full_tune_out} for comparison). We find the two values agree within \(200\)~MHz, and hence use this as our uncertainty for the window birefringence. 
 
\subsubsection{Polarization across the beam}
We have also observed a small shift in the probe beam polarization with the location of the measurement in the beam. We have identified the mirrors in our probe beam delivery system to be responsible. To characterize this we have probed the polarization at many points across the beam, repeating a similar process where the \(f_{\mathrm{TO}}(-1,0)\) is found from a single polarization measurement location. We have found that variation in the value of \(f_{\mathrm{TO}}(-1,0)\) is up to \(400\)~MHz away from the beam center. However, we must note that the contribution of these polarizations to the total value is weighted by their power in our measurement. Accordingly the power weighted values give an uncertainty of \(150\)~MHz for the central measurement of polarization.

\subsection{Linearity} \label{sec:syst.subsec:lin}
The first stage in the measurement of \(f_{\mathrm{TO}}(-1,0)\) assumes a linear response of the  perturbing trap frequency with respect to the optical frequency. This is, however, only an approximation as the polarizability itself is nonlinear with optical frequency at a large enough scale ($\sim$THz). Further, the relation $\Omega_{\text{net}}^2=\Omega_{\text{trap}}^2+\Omega_{\text{probe}}^2$ assumes that the probe beam is harmonic, which is only approximately true for a Gaussian beam.

\newcommand{\subfigwidth}{0.28\linewidth}
\begin{figure}
    \centering
    \setlength{\FrameRule}{3pt}
    \setlength{\FrameSep}{0pt}
    \begin{framed}
    \large{After chamber}\\
    \subfigure[]{\includegraphics[width=\subfigwidth]{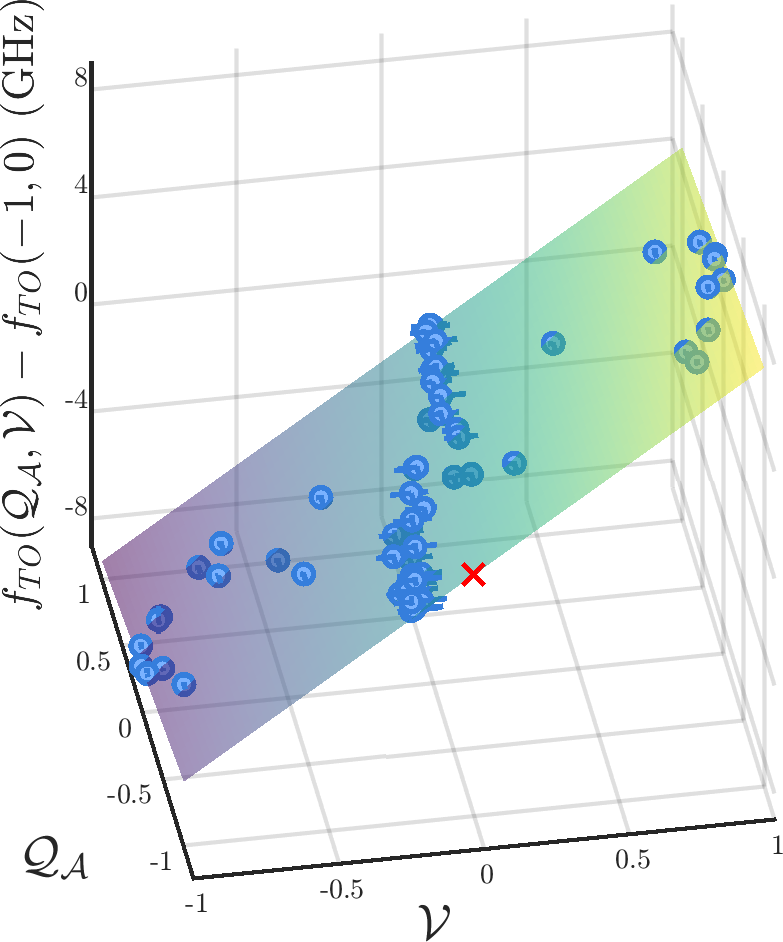}}
    \hspace{0.5cm}%
    \subfigure[]{\includegraphics[width=\subfigwidth]{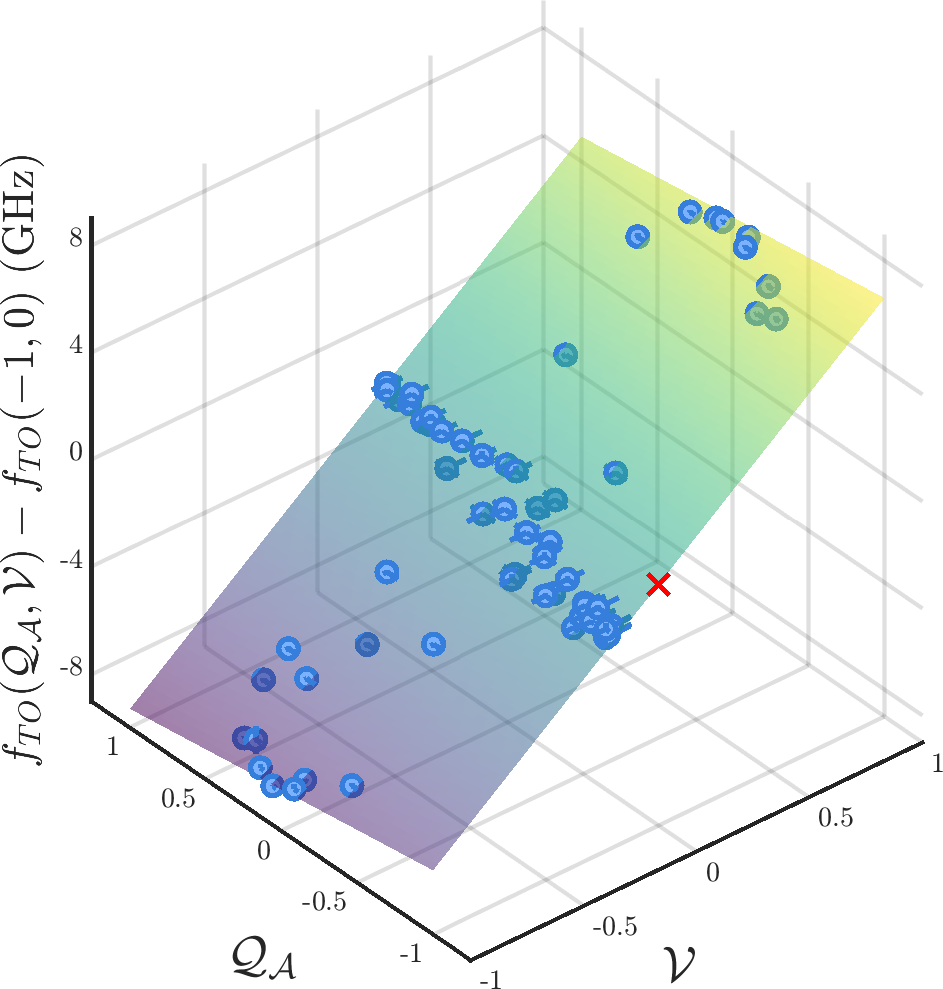}}
    \hspace{0.5cm}%
    \subfigure[]{\includegraphics[width=\subfigwidth]{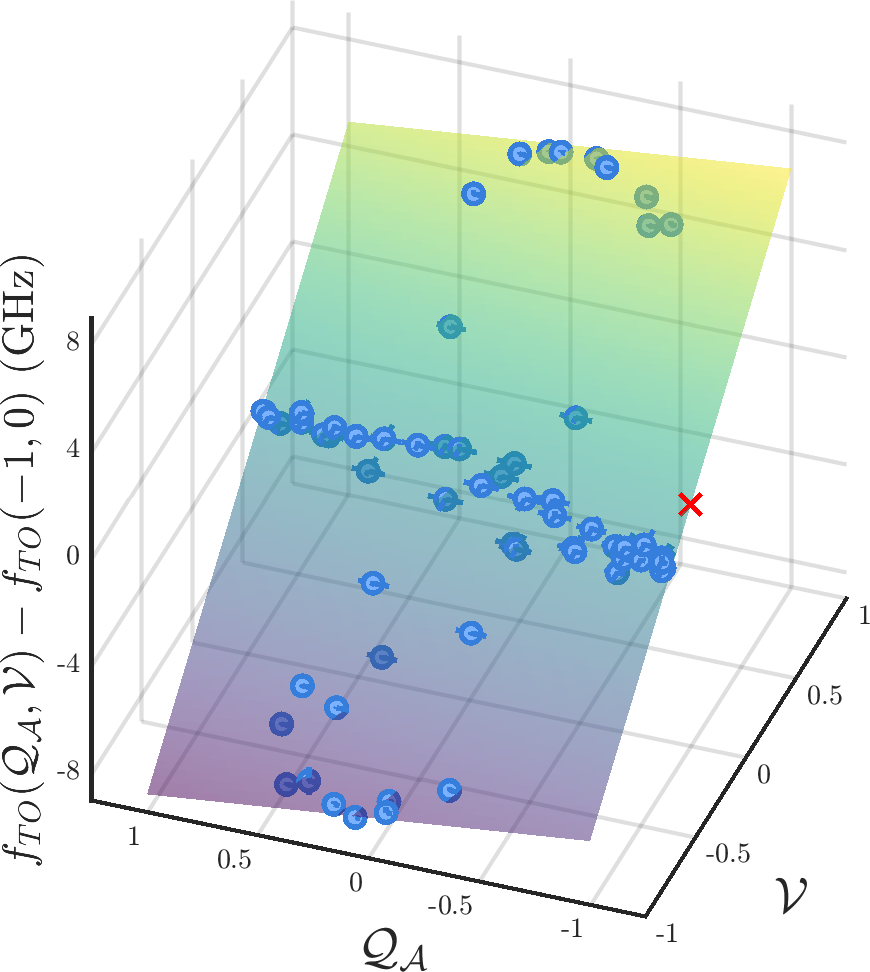}}
  \end{framed}
  \vspace{-0.8cm}
    \begin{framed}
    \large{Before chamber}\\
    \subfigure[]{\includegraphics[width=\subfigwidth]{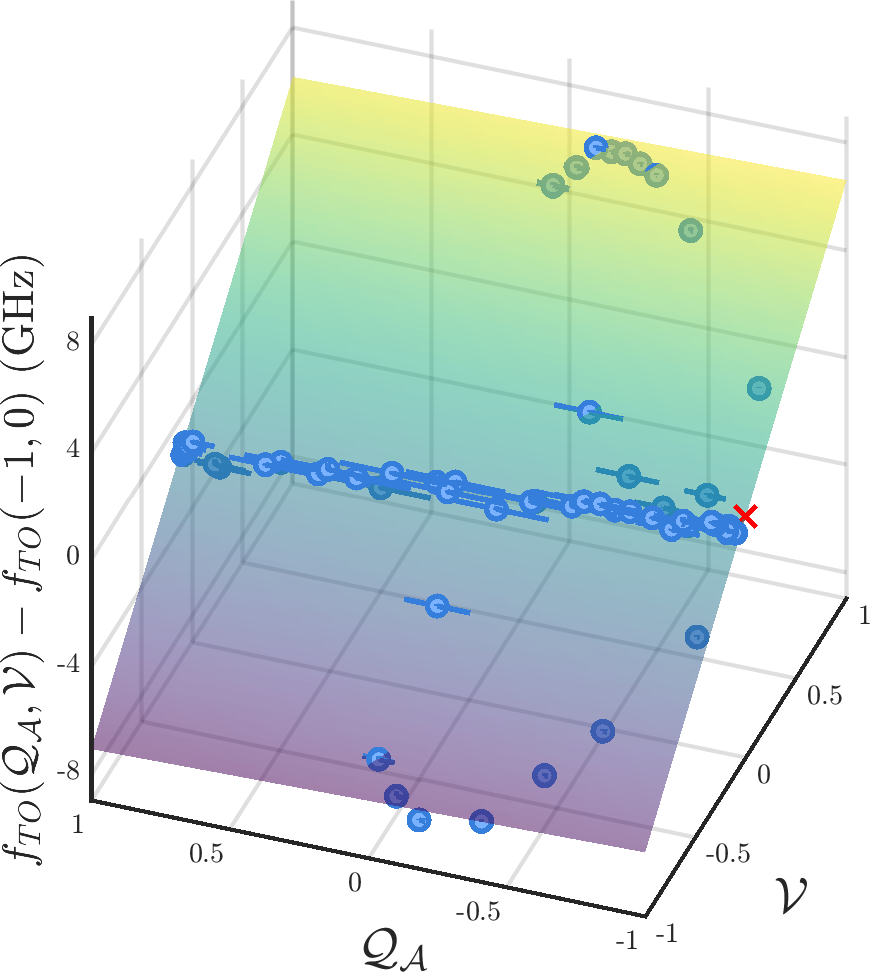}}
    \hspace{0.5cm}%
    \subfigure[]{\includegraphics[width=\subfigwidth]{figs/rev2/to_vq_dependence_2_pre}}
    \hspace{0.5cm}%
    \subfigure[]{\includegraphics[width=\subfigwidth]{figs/rev2/to_vq_dependence_2_pre}}
  \end{framed}
\caption{Visualization of the fit to the measured tune-out $f_{TO}$ as a function of the $\mathcal{Q_{A}}$,\, $\mathcal{V}$ polarization parameters (see Eq.~(\ref{eqn:tune_out_eq}) for full functional form). The top set (a-c) of plots shows the fit using the polarization data taken after the vacuum chamber, while the bottom set (d-f) uses data taken before the vacuum chamber.  Each blue (round) point represents a tune-out measurement with a given polarization state of the probe beam. 
Horizontal error bars represent uncertainty in the light polarization measurement.
Note that (a-c) and (d-f) are different rotational views for the same data sets respectively. 
The red cross shows the resulting value for \(f_{\mathrm{TO}}(-1,0)\), which is \(f_{\mathrm{TO}}(-1,0)=725\, 736\, 810(40)\)~MHz for the top data, and \(f_{\mathrm{TO}}(-1,0)=725\, 736\, 610(40)\)~MHz for the bottom data. The other fit parameters are \(\beta^V \cos(\theta_k)=13240(70)\)~MHz, and \(\beta^T \sin^2(\theta_k)=1140(20)\)~MHz.}
\label{fig:full_tune_out}
\end{figure}

\subsubsection{Method Linearity}
\begin{figure}[b]
    \centering
    \includegraphics[width=0.65\textwidth]{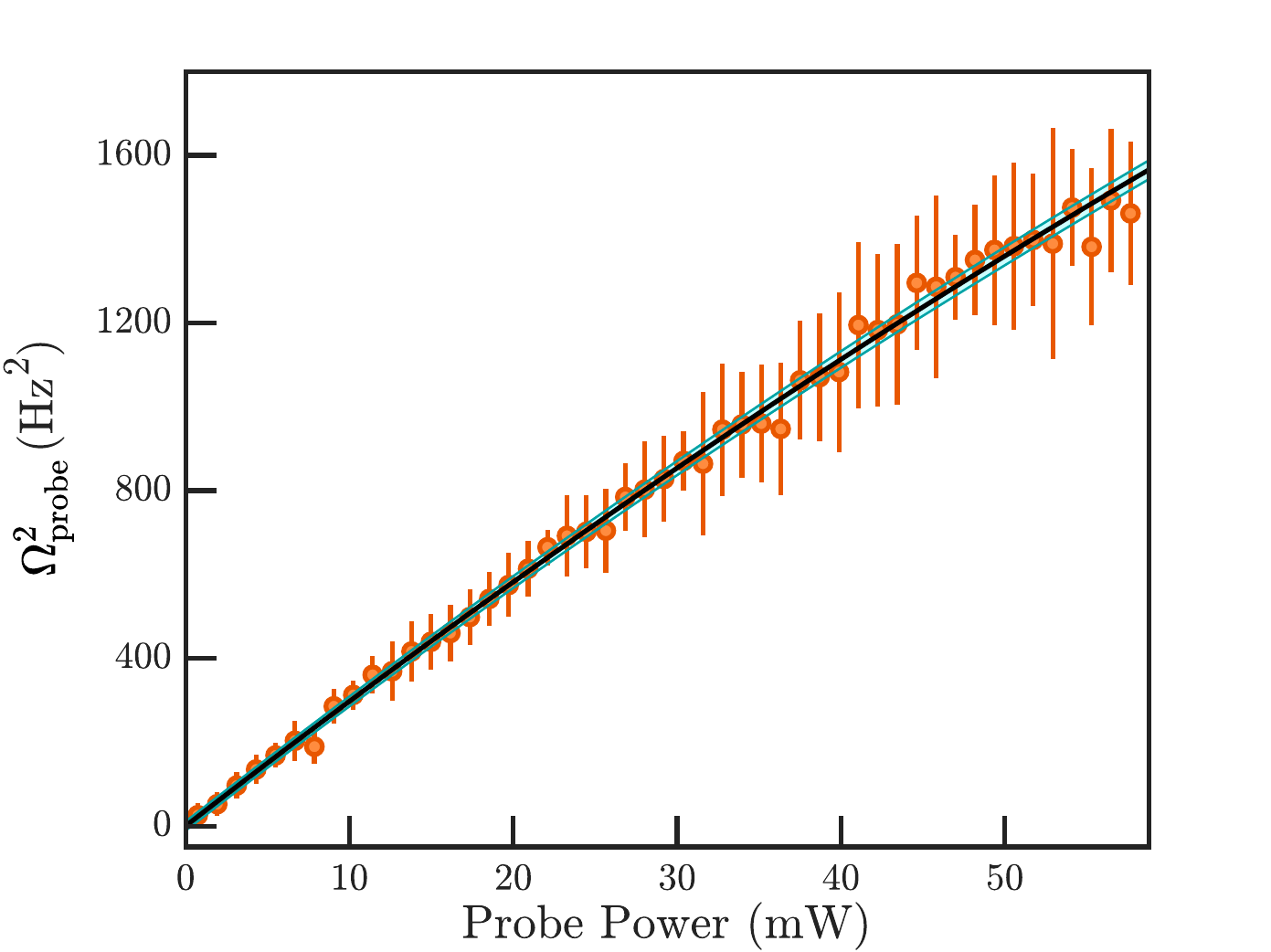}
    \caption{Linearity of the derived probe beam trap frequency with probe beam power. The probe beam optical frequency is set 20~GHz to the blue of the tune-out to produce a strong potential. The fit is found to be \( \Omega_{\mathrm{probe}}^2 = a  P - b P^2 \), with $\chi^2$/dof=0.992, where P is the probe beam power (in watts) and \(a\), \(b\) are the fit parameters with values \(a=\text{30.3(1)}\times 10^{-3}\text{~Hz}^2\text{W}^{-1}\), \(b= \text{0.06(2)} \times 10^{-6}\text{~Hz}^2\text{W}^{-2}\). Higher order terms in the fit function were not found to be statistically significant. The shaded region shows the fit's \(1\sigma\) confidence interval.
    }
    \label{fig:probe_beam_linearity}
\end{figure}

To estimate the shift from the potential measurement method we study the derived \(\Omega_{\mathrm{probe}}\) as a function of probe beam power at a fixed polarization and detuning from the tune-out (see Fig.~\ref{fig:probe_beam_linearity}). We find that a second order model of the response is sufficient to describe the behaviour. From this dependence we find that the tune-out may be shifted by \(-24(30)\)~MHz.

\subsubsection{Polarizability Linearity}
The polarizability $\alpha(f)$ about the tune-out is approximately linear in frequency. However, this is not exact. Using a model of the polarizability we find that fitting a linear dependence over 4~GHz either side of the tune-out results in a fit intercept which is $-88(1)$~kHz from the true tune-out. This shift would increase to $-9.6(2)$~MHz if measurements were taken 40~GHz about the tune-out. A quadratic fit reduces this shift to 0.6~kHz and 40~kHz in the 4~GHz and 40~GHz cases respectively.

\begin{figure}[t]
    \centering
    \includegraphics[width=0.85\textwidth]{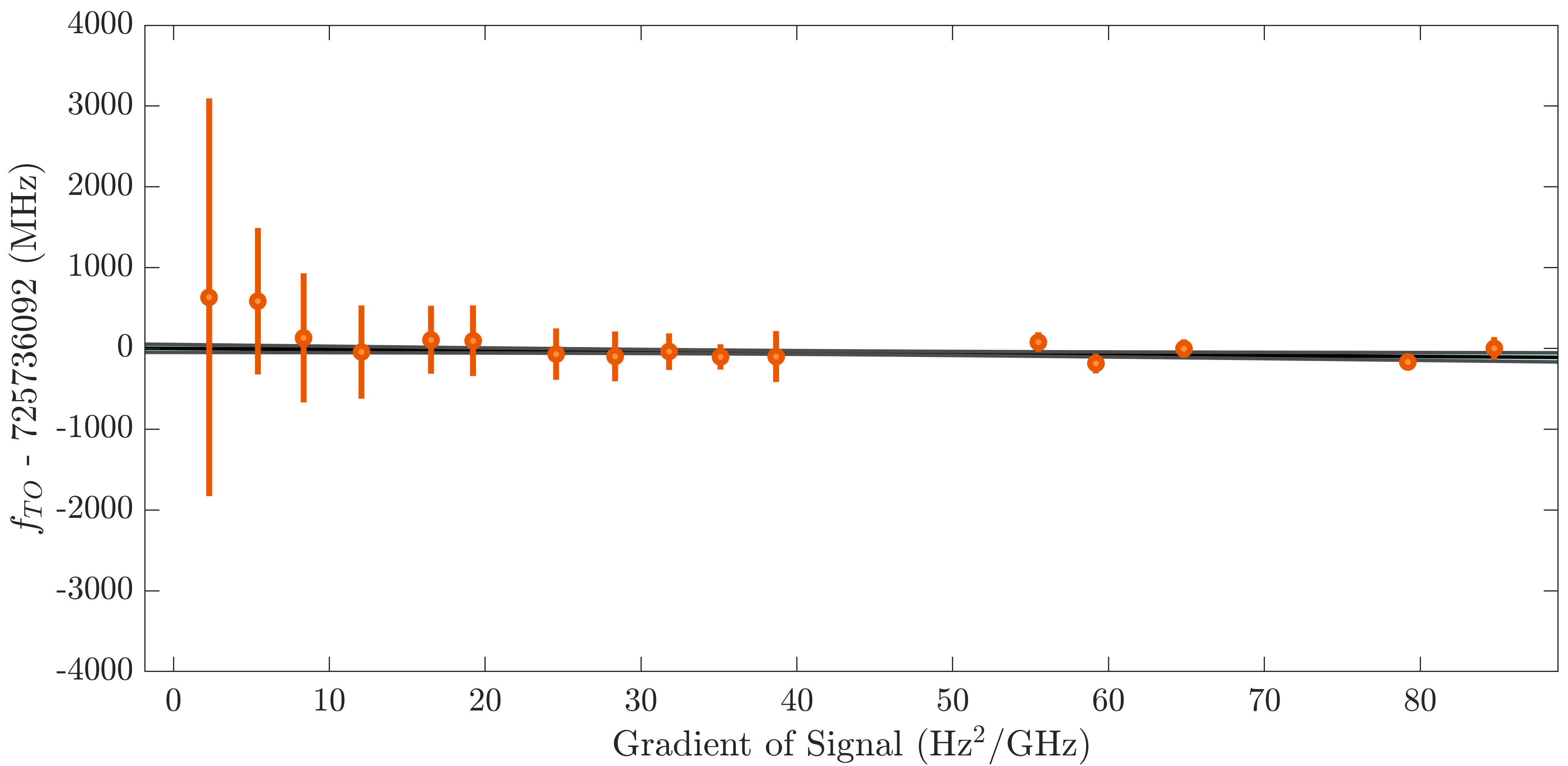}
    \caption{Measured tune-out dependence on probe beam intensity. The highest signal gradient (right) corresponds to a peak light field intensity of \(\sim4\times10^{8}\: \mathrm{W}\cdot \mathrm{m}^{-2}\).
    An uncertainty-weighted linear fit to the  data set with parameters offset\(=725\,736\,083(50)\)~MHz and gradient -1.2(1.5) MHz/(Hz\(^2\)/GHz), along with shaded region corresponding to 1$\sigma$ confidence interval, is also shown. 
    This fit has $\chi^2$/dof=0.987 and determines that the gradient dependent tune-out shift, which is within error of no effect for experimentally relevant intensities, $30(50)$~MHz for the power used in the main measurement. 
    }
    \label{fig:hyperpolarizability}
\end{figure}

\subsection{Validating linearity against theory}
 
We can use the values of the fit parameters to get an idea how large the non-frequency, i.e. low accuracy, contribution to the measurement are. That is, we can make a comparison, albeit with low precision, between theoretical predictions for the coefficients $\beta^V$ and $\beta^T$, and also compare the predicted and experimental value of the polarizability gradient $\partial \alpha/\partial f$. Also importantly, from these values we can obtain an idea of the size of the non-frequency, \textit{i.e.} low accuracy polarization, contributions to the measurement. These values are summarized in Table \ref{tab:param_compare} and the means by which we estimate them are given below. These values are not the primary interest of this work and accordingly we do not expect precise agreement between theory and experiment. However, we should expect that the range of values obtained from the experiment should at least include the theoretical prediction.

	\begin{table}[t]
	    \centering
	    \begin{tabular}{l c c c c c}
	        \hline\hline
	         Quantity & $\partial \alpha/\partial f$ (C m$^2$ V$^{-1}$ Hz$^{-1}$) && $\beta^V$ (MHz) &&  $\beta^T$ (MHz)\\
	         \hline
	         Experiment & $5(7)\times10^{-54}$ &&  $1.5(1)\times10^4$ && $2.6^{+1.2}_{-0.6}\times10^3$\\
	         Theory & $1.78\times10^{-53}$ && $1.22\times10^4$ && $3.4\times10^3$\\
	         Ratio exp./thr. & 0.3(4) && 1.2(1) && \(0.8^{+0.4}_{-0.2}\)\\
	         \hline\hline
	    \end{tabular}
	    \caption{Comparison between the theoretical and experimentally-determined values for the gradient of the polarizability, and the reduced-vector and reduced-tensor polarizabilities. Note that the latter two are coupled via the fit model (through $\theta_k$) and hence this determination is not unique. Nonetheless, these comparisons show that the experimental values are generally consistent with the theoretical predictions. The parentheses denote the 1$\sigma$ uncertainty in the final digit. Uneven uncertainty intervals are quantified by the upper (superscript) and lower (subscript) values. All experimental values are within 2$\sigma$ of the predicted values.}
	    \label{tab:param_compare}
	\end{table}

	We note that at the tune-out the polarizability is zero and thus its value near $f_{TO}$ is determined by the gradient $\partial \alpha/\partial f$.  Hence, here we provide a comparison between theoretical and experimental values for $\partial\alpha/\partial f$, $\beta^V$, and $\beta^T$, which completely characterize the polarizability in relevant conditions.

	We can estimate the gradient $d\alpha/df$ by computing, from Eq. (\ref{eqn:response_eq}),
	\begin{align}
	\frac{\partial~ Re(\alpha)}{\partial~f}&=A^{-1}\frac{d~\Omega_{\text{probe}}^2}{d~f},~\textrm{where} \label{eqn:grad_vs_intesnity}\\
	    A&=\frac{P}{m \epsilon_{0} c \pi^3  w_0^4}=\frac{1}{ \pi^2 m \epsilon_{0} c }\frac{I}{2 w_0^2}.
	\end{align}
	Fig.~2C in the main article displays a response with a slope of approximately $d\Omega^2/df \approx 30$ Hz$^2$/GHz, which is typical of the data collection runs. 
	We have records of the probe beam power and profile obtained from measurements outside the vacuum chamber. 
	However, the vacuum window (through which the beam passes) could subtly alter the beam profile and focal point, and a beam this tightly focused will have non-Gaussian aberrations arising from imperfections in the optics. 
	Hence we adopt a conservative margin of error for the beam waist, using the range 15(5) $\mu$m. 
	Similarly we assume knowledge of the beam power at the trap to within about 25$\%$, i.e. $P\approx140(30ou)$ mW. The conversion factor consistent with these values is $A\approx5(7)\times10^{45}$ kg A$^{-2}$  s$^{-6}$. 
	Thus the experimental value of the polarizability gradient is 5(7)$\times10^{-54}$ C m$^2$ V$^{-1}$ Hz$^{-1}$, where the bracketed digit includes the statistical uncertainty across all runs used in the analysis and the (dominant) uncertainty due to imperfect knowledge of the beam parameters at the interrogation zone. 
	The uncertainty interval is shown in Table \ref{tab:param_compare} and is broadly consistent with the theoretical value.
	The accuracy of this estimate is fundamentally limited by the quartic dependence on the beam waist, which was not precisely quantified as it is not a critical parameter (Eq.~(\ref{eqn:response_eq}) shows that the waist must only be stable, not precisely measured, to determine the tune-out frequency). However, as we discuss below in the context of the hyperpolarizability, the error induced by imperfect knowledge of the probe beam intensity is not significant.

	Our nonlinear fitting procedure returns $\beta^T\sin^2(\theta_k)=1140(20)$~MHz but, as noted above, unique values of $\beta^T$ and $\theta_k$ are not obtainable because the physical model is over-articulated (i.e. has more parameters than degrees of freedom due to physical constraints). 
	From simulations of our magnetic trap we can estimate the angle $\theta_k\approx 30^\circ$, which is consistent with the fitted value of $25^\circ$. Working with an estimate of $\theta_k = 27.5\pm(5)^\circ$, we determine the values shown in Table \ref{tab:param_compare}. These calculations indicate effect sizes that are of the same order of magnitude as theoretical predictions, and whose uncertainty bounds also include the predicted value.
	Similarly, the theoretical value for $\beta^V$ is similar to the experimental value.

\subsection{Hyperpolarizability}
The above work assumes that the energy shift in the ground state due to a nonzero polarizability is linear with light field intensity. To reveal the error from this approximation we use a combination of theoretical predictions and experimental measurements.
\subsubsection{Theoretical treatment}
The energy of state in an atomic system perturbed by an electric field \(E\) with frequency \(f\) is given by 
\begin{equation}
\mathcal{E}=\mathcal{E}_0 - \frac{1}{2} \alpha(f) E^2 - \frac{1}{24} \gamma(f) E^4 + \ldots \, , \label{eqn:energy_shift}
\end{equation}
where \(\mathcal{E}_0\) is the unperturbed energy, \(\alpha(f)\) is the dynamic polarizability as defined above, and \(\gamma(f)\) is the frequency dependent second hyperpolarizability.  For monochromatic optical radiation the time averages electric field amplitude can be found from the intensity (I),
\begin{equation}
    E^2=\frac{2 I}{c \epsilon_0},
\end{equation}
where \(c\) is the speed of light and \(\epsilon_0\) is the permittivity of free space. A measurement of the tune-out will be shifted by an amount such that the dynamic polarizability cancels the contribution from the hyperpolarizability. Using the fact that \(\mathcal{E}=\mathcal{E}_0\) at the measured tune-out by definition, and taking a first order Taylor expansion of the polarizability about the tune-out, Eq.~(\ref{eqn:energy_shift}) gives 
\begin{equation}
 (f_\mathrm{TO}-f_{\mathrm{TO},\alpha} ) = - \frac{1}{12} \gamma(f) \left(\frac{2 I}{c \epsilon_0}\right) \left( \frac{\partial\alpha}{\partial f}\bigg|_{f=f_\mathrm{TO}}  \right)^{-1}. \label{eqn:hyp_pol_shift}
\end{equation}
where \(f_\mathrm{TO}\) is the measured tune-out frequency and \(f_{\mathrm{TO},\alpha}\) is the tune-out frequency without any hyperpolarizability correction. From our theoretical calculations, the dynamic hyperpolarizability at the tune-out is \(6.964\times10^{-58}\: \mathrm{C}^4\mathrm{m}^4\mathrm{J}^{-3}\) (about $-1.12\times10^{7}$ atomic units). The probe beam intensity used in the present experiment is less than $10^{9}\: \mathrm{W} \mathrm{m}^{-2}$, hence the magnitude shift due to the hyperpolarizability is less than 1.5~MHz, well below other systematic errors.

\subsubsection{Experimental treatment}
As an independent determination of the above theoretical treatment we study the dependence of the measured tune-out on the light intensity. The gradient of the probe beam trap frequency \(\Omega_{\text{probe}}^2\) with respect to optical frequency provides an indirect measurement of the intensity at the position of the atoms. As a proxy for the intensity it accounts, to first order, for changes due to probe beam misalignment. To this end we measure the tune-out at a variety of probe beam intensities (see Fig.~\ref{fig:hyperpolarizability}) and find no effect within error (a shift of $30(50)$~MHz for the probe beam intensity used in the main measurement).

\subsubsection{Comparison of methods}
In order to directly compare our experimental results regarding the hyperpolarizability with the theoretical expectations we need to relate the intensity dependence of Eq. (\ref{eqn:hyp_pol_shift}) to the signal gradient dependence shown in Fig.~\ref{fig:hyperpolarizability}. This can be done by using Eq. (\ref{eqn:grad_vs_intesnity}) (based on Eq. (\ref{eqn:response_eq})) to write the signal gradient \(\frac{d \Omega_{\text{probe}}^2 }{d f}\) in terms of the peak intensity \(I_0\),
\begin{align}
    \frac{d \Omega_{\text{probe}}^2 }{d f} &= \frac{1}{2\pi^2m w_{0}^{2} c \epsilon_0} \frac{\partial \alpha}{\partial f} I_0. \label{eqn:signal_grad_dep}
\end{align}
As the polarizability gradient \(\frac{\partial \alpha}{\partial f}\) is independent of intensity, the signal gradient is directly proportional to the peak intensity. If we approximate the average intensity the atoms experience using the peak intensity, \(I \approx I_0\), we can then rearrange Eq. (\ref{eqn:signal_grad_dep}) and substitute into Eq. (\ref{eqn:hyp_pol_shift}) to obtain the expected hyperpolarizability dependence on signal gradient,
\begin{align}
 I &= 2\pi^2m w_{0}^{2} c \epsilon_0 \left(\frac{\partial \alpha}{\partial f}\right)^{-1} \frac{d \Omega_{\text{probe}}^2 }{d f} \\
 f_\mathrm{TO} &= f_{\mathrm{TO},\alpha} - \frac{\gamma(f) }{12} \left(4\pi^2m w_{0}^{2}\right) \left( \frac{\partial\alpha}{\partial f}\bigg|_{f=f_\mathrm{TO}}  \right)^{-2} \frac{d \Omega_{\text{probe}}^2 }{d f}.
\end{align}

As we are interested specifically in the gradient of the tune-out frequency with the signal slope, we have,
\begin{align}
-\frac{\gamma(f) }{12} \left(4\pi^2m w_{0}^{2}\right) \left( \frac{\partial\alpha}{\partial f}\bigg|_{f=f_\mathrm{TO}}  \right)^{-2}.
\end{align}
Given theoretical values of the second hyperpolarizability \(\gamma(f_\mathrm{TO}) = 6.964 \times 10^{-58} \, \text{C}^4 \text{m}^4 \text{J}^{-3}\), the mass of helium \(6.646477 \times 10^{-27}\)~kg, beam waist \(w_0 = 10\)~\(\mu\)m, and \(\frac{\partial \alpha}{\partial f} = 1.78 \times 10^{-53} \, \text{C} \text{m}^2 \text{V}^{-1} \text{s}\) (as discussed above), the value of the hyperpolarizability dependence is \(-1.1(7) \times 10^{13}=-0.011(7)\)~MHz/(Hz\(^2\)/GHz), which is consistent with the experimental value of \(-1(2) \times 10^{15}=-1(2)\)~MHz/(Hz\(^2\)/GHz) (see Fig.~S6), \textit{i.e.} no observed effect within error.

\subsection{Broadband Light}
The (superlinearly) increasing atomic polarizability with detuning from the tune-out puts demanding constraints on the spectral purity of the laser used for this measurement. Broadband light can be produced by amplified spontaneous emission of the laser which has been deleterious for measurements in other species \cite{HolmgrenPhd}. It is the product of the spectral power distribution and the polarizability (which heavily weights the tails) to produces a potential and shifts the apparent tune-out from its true value. To minimize this error we use a frequency doubled laser system which provides some suppression of the background through the doubling cavity. 

Suppression is then further improved with a series of optical filters. The first filter is a 450~nm shortpass filter (Thorlabs FESH0450) with an optical density of $>5$ between 450~nm and 1200~nm.  The second is a 415~nm band-pass filter (Semrock FF01-415/10-25) with a FWHM of 27~THZ (15.3~nm) and an optical density of $>4$ out of band between 250-399~nm \& 431-1100~nm. The final filter is an angle-tunable filter with a FWHM of 0.9~THz (0.5~nm), which we center on \(\sim 413\)~nm.

\begin{figure}[t]
    \centering
    \includegraphics[width=0.85\textwidth]{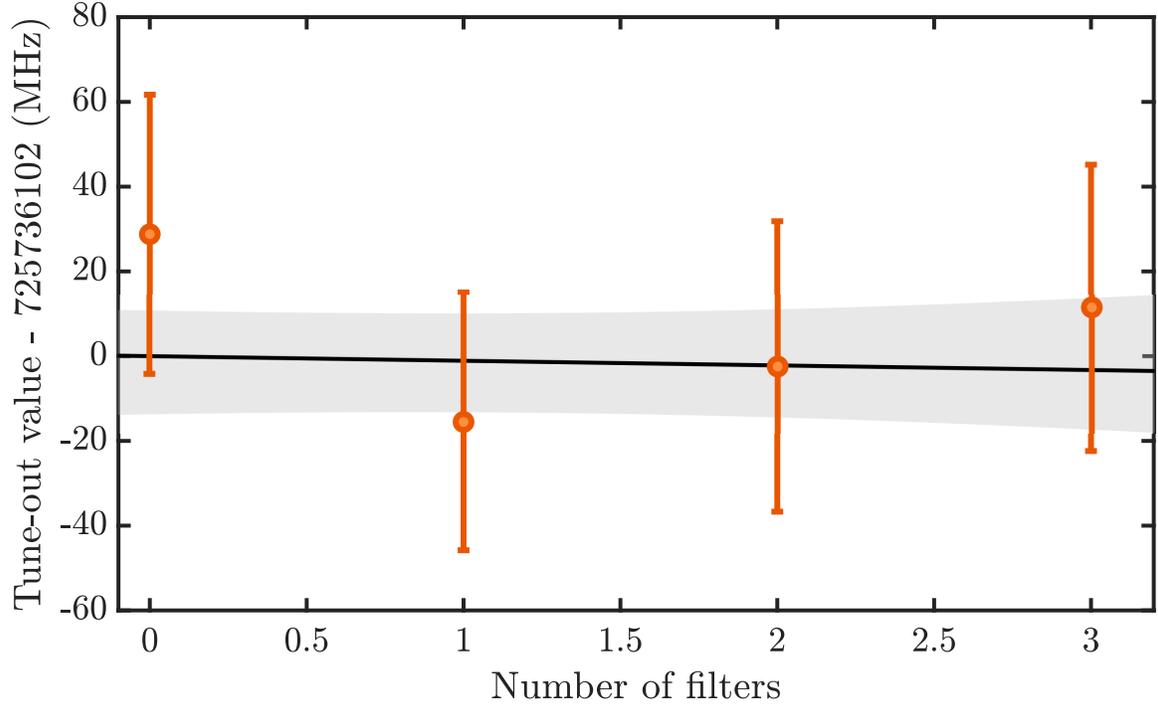}
    \caption{Measured tune-out frequency for a constant probe beam polarization as a function of the number of filters the probe beam light passes through. We find that the gradient of this dependence is -1(5) MHz/filter, with a fit $\chi^2$/dof=0.6406, and hence zero within error.
    }
    \label{fig:broadband light dependence}
\end{figure}

To estimate the remaining error from spectral background one could in principle measure the power spectral density using a spectrometer however the dynamic range needed to see these small powers near the main laser frequency makes this direct approach unfeasible. We therefore employ a scheme similar that used in \cite{PhysRevA.92.052501} measuring the tune-out as a function of the number of filters the probe beam light passes through with progressively narrow spectral filtering. This dependence used to estimate  the final measurement shift. From our experimental fit (see Fig.~\ref{fig:broadband light dependence}), we find that within error there the measured tune-out frequency is independent of the amount of light filtering. We find a standard deviation between the various filters of \(30\)~MHz, and thus take this as our uncertainty in the broadband light shift.
 
\subsection{DC Electric Field}
To give the shift in the tune-out from background dc electric fields, we may use a similar approach to the hyperpolarizability above. A worst case estimate of our electric field is $2 \, \,\text{kV}\cdot\text{m}^{-1}$. This results in a worst-case shift of \(10^{-2}\)~MHz.

\subsection{Mean-Field Shift}

In traditional transition frequency measurements \cite{Rengelink2018} it is common to consider the \textit{cold collisional shift} which arises from the difference in mean-field interaction energy between the ground and excited state. 
In a measurement of the tune-out this mechanism affects each of the constituent transitions from the ground state, with the net shift in the tune-out weighted by the relative contribution of each transition.
The scale of a mean-field transition shift is generally far smaller than the uncertainty of our tune-out measurement. 
For example, an in-trap measurement of the singlet-triplet transition in helium reported a maximum expected shift of 4.4~kHz  \cite{Rengelink2018} (for a comparable density). 

For bosons with sufficiently low temperatures such that \textit{s}-wave collisions dominate
(less than 100~mK for He*)
then the density dependent local energy level shift $\Delta E_p$ of a state $p$ is given by
\cite{PhysRevLett.81.3807}
 \begin{align}
    \label{ch:to.eq:mean_field_shift}
     \Delta E_p = \frac{8 \pi \hbar^2}{m} (a_{pq} n_q + a_{pp} n_p),
 \end{align}
where $a_{pq}$ is the scattering length between atoms in the $p$ and $q$ states, and $n_p$ is the density of the atoms in state $p$. In this treatment we use the negligible excitation probability of excited states when interrogating the tune-out to ignore the contributions from non-elastic collisions. The frequency shift $\Delta_{mf} f(p,q)$ of a transition $p - q$ can be found by combining the energy change in the lower ($p$) and upper ($q$) states as:
\begin{align}
\Delta_{mf} f(p,q) =& \frac{8 \pi \hbar}{m} \left(
 -a_{pp} n_p 
 -a_{pq} n_q  
 +a_{qq} n_q 
 +a_{pq} n_p 
 \right)\\
 \approx& \frac{8 \pi \hbar}{m} n_p\left(
 -a_{pp}  
 +a_{pq} 
 \right).
 \end{align}
Where in the second line we once again assume negligible excitation probability such that $n_q\approx 0$. 
In this treatment we have assumed that the energy level shift dominates the effect on the tune-out over any mean-field induced change in the oscillator strengths. The total shift in the tune-out from the mean-field shift $\Delta_{mf}f_{\mathrm{TO}}$ is found by weighting the shift of each transition by the partial derivative of the tune-out with respect to that transition frequency, the dominant terms are found numerically to be
\begin{align}
\Delta_{mf}f_{\mathrm{TO}} \approx &
0.4916 \ \Delta_{mf} f(2^{3}S_{1},3^{3}P_{2}) + 
0.2951 \ \Delta_{mf} f(2^{3}S_{1},3^{3}p_{1}) +
\nonumber
\\
& 
0.0984 \ \Delta_{mf} f(2^{3}S_{1},3^{3}P_{0}) +
0.0327 \ \Delta_{mf} f(2^{3}S_{1},2^{3}P_{2}) 
\end{align}
where $\Delta_{mf} f(p,q)$ is the mean-field shift of the $p - q$ transition.

The scattering length of the \(\MetastableState-\MetastableState\) state collisions is of the order of \(10\)~nm \cite{PhysRevLett.96.023203} but unfortunately the scattering lengths between the metastable state and the relevant excited states have not been calculated or empirically measured (a similar situation to \cite{vanRooij196}).
This might encourage theoretical studies to predict the inter-state scattering lengths for helium as a resource for future in-trap spectroscopic investigations.
To set a strong upper bound on the magnitude of the mean-field shift we consider the extreme case where the scattering length between the metastable state and the excited states is 300~nm. 
This scattering length is comparable to that found near a strong Feshbach resonance in other species \cite{PhysRevLett.112.190401} and is far greater than any known scattering length in helium. For the typical experimental conditions in this work the mean density of the BEC is $\approx 9\cdot10^{18}$~m$^{-3}$ using the Thomas-Fermi approximation.
This combination of density and excited state scattering length would produce a mean-field shift in the tune-out of $\Delta_{mf}f_{\mathrm{TO}}\approx1$~MHz which we take as an upper bound for the mean-field shift magnitude.

\section{Theory and Results for \(\TO\) Tune-Out Frequency}
 \label{ch:sm.sec:theory}
The interaction of an isolated atom with a freely propagating laser field
is more properly regarded as a zero in the coherent Rayleigh scattering cross section, rather
than as a zero in the dynamic polarizability as is the case for an atom in an optical lattice (see Refs.~\cite{PhysRevLett.124.203201,PhysRevA.92.052501,PhysRevA.99.040502}).
  The two formulations are identical to the lowest order, but not when retardation (finite wavelength)
effects are taken into account, as further discussed in this section.  For convenience,
we will continue to use the $\alpha_{\rm d}(\omega)$ notation to denote a zero in the
Rayleigh scattering cross section.

Table \ref{tab:theory} provides a summary of the main theoretical contributions to the \(\TO\) tune-out frequency in helium, calculated in a nonreltivistic QED (nr-QED) approach. The nr-QED method is briefly described in the following subsection, followed by the results.\\[\baselineskip].

\subsection{The nr-QED Method}
The nr-QED method starts with the nonrelativistic Schr\"odinger equation, and then includes the terms in the Pauli form of the Breit interaction by perturbation theory \cite{BetheSalpeter}.  Our nr-QED formalism is similar to that presented in Refs.~\cite{Piszczatowski2015,Puchalski2016,PhysRevA.101.022505}.  The basic expression for the dynamic polarizability is of second-order in the interaction with the external electromagnetic field of frequency $\omega$, and the additional contributions from relativistic and QED effects are included by means of an additional perturbation.  For brevity, define the resolvent operator to be,
\begin{equation}
{\cal R}(\omega) = Q(H_0 - E_0 + \hbar\omega)^{-1}Q,
\end{equation}
 where $H_0$ is the field-free Hamiltonian, $E_0$ is the unperturbed energy of the $2^3S_1$ state, and $Q = 1 - |\psi_0\rangle\langle \psi_0|$ is a projection operator.  In the electric dipole approximation, the frequency-dependent dipole polarizability is then defined by the symmetric combination
  $\bar{\alpha}_{\rm d}(\omega) = \frac12[\alpha_{\rm d}(\omega) + \alpha_{\rm d}(-\omega)]$, where
\begin{equation}
\alpha_{\rm d}(\omega) = 2\langle\psi_0|\edotr{\cal R}(\omega)\edotr|\psi_0\rangle,
\end{equation}
$\mbox{\boldmath{$\hat{\rm e}$}}$ is a polarization vector pointing in the direction of the electric field, and ${\bf r} = {\bf r}_1 + {\bf r}_2$.  Any additional time-independent perturbation $\hat{X}$ can then be included according to the $(2+1)$-order double perturbation expression,
\begin{equation}
\delta\alpha_{\rm d}^{\hat{X}}(\omega) = 2\langle\psi_0|[2\edotr{\cal R}(\omega)\edotr{\cal R}(\omega)\hat{X}
+ \edotr{\cal R}(\omega)(\hat{X} - \langle\hat{X}\rangle){\cal R}(\omega)\edotr]|\psi_0\rangle
\end{equation}

The tune-out frequency $\omega_{\rm TO}$ corresponds to the condition (for $S$-states)
\begin{equation}
\label{eq:baralpha}
|{\bf E}|^2\left[\bar{\alpha}_{\rm d}(\omega) + \sum_{\hat{X}} \delta\bar{\alpha}_{\rm d}^{\hat{X}}(\omega) +
\delta\bar{\alpha}_{\rm d}^{\partial_{\cal E}^2\ln k_0}(\omega) + \delta\alpha_{\rm d,ret}(\omega)\right]
+ |{\rm B}|^2\left[\chi + \delta\chi_{\rm ret}(\omega)\right]= 0
\end{equation}
for the Rayleigh scattering amplitude to vanish \cite{PhysRevA.99.041803}, where $\chi$ is the magnetic susceptibility, the sum over $\hat{X}$ runs over all the perturbations included in the calculation, and $\delta\bar{\alpha}_{\rm d}^{\partial_{\cal E}^2\ln k_0}(\omega)$ is an additional QED correction due to the field dependence of the Bethe logarithm discussed further below, together with the retardation corrections $\delta\alpha_{\rm d,ret}(\omega)$ and   $\delta\chi_{\rm ret}(\omega)$.  The electric and magnetic fields {\bf E} and {\rm B} are related to the photon vector potential ${\rm A} = A_0\hat{\bf e}
e^{i{\bf k}\cdot{\bf r} - i\omega t}$ by $E = -(1/c)\partial{\bf A}/\partial t$ and ${\bf B} = \nabla\times{\bf A}$, and so both $|{\rm E}|^2$ and $|{\rm B}|^2$are proportional to $\omega^2$.

For the relativistic corrections, the operators $\hat{X}$ in Eq.\ (\ref{eq:baralpha}) consist of the spin-independent terms $H_1 = -(p_1^4 + p_2^4)/(8m^3c^2)$, the orbit-orbit interaction
\begin{equation}
H_2 = -\frac{e^2}{2(mc)^2r_{12}}\left[{\bf p}_1\cdot{\bf p}_2 + \frac{{\bf r}_{12}\cdot({\bf r}_{12}\cdot{\bf p}_1){\bf p}_2}{r_{12}^2}\right],
\end{equation}
and the Darwin term $H_4 = \pi(\alpha e a_0)^2[Z\delta^3({\bf r}_1) - \delta^3({\bf r}_{12})]$, together with the  spin-dependent spin-spin interaction
\begin{equation}
H_5 = \frac{4\mu_0^2}{r_{12}^3}\left[{\bf s}_1\cdot{\bf s}_2 - \frac{3({\bf s}_{1}\cdot{\bf r}_{12})({\bf s}_2\cdot{\bf r}_{12})}{r_{12}^2}\right],
\end{equation}
where ${\bf r}_{12} = {\bf r}_1 - {\bf r}_2$ is the interelectron coordinate, $a_0$ is the Bohr radius, $\mu_0
= e\hbar/(2mc)(1 + \alpha/(2\pi) + \cdots)$ is the Bohr magneton, and $\alpha\simeq1/137.03599906$ is the fine structure constant. The spin-orbit and spin-other-orbit terms that are normally part of the Breit interaction do not contribute in lowest order after summing over fine structure for the intermediate $^3P_J$ states. Also, the $\delta^3({\bf r}_{12})$ term does not contribute for triplet states. 

\begin{table}[t]
\caption{\label{tab:theory}Summary of theoretical contributions to the helium $\TO$ manifold tune-out frequency near 725.7 THz.}
\begin{ruledtabular}
\begin{tabular}{l r r}
Quantity    &Value (MHz) & Uncertainty (MHz) \\
\hline
\multicolumn{3}{c}{\textbf{Nonrelativistic and Relativistic terms}} \\
Nonrelativistic (NR) & 725\,645\,115& 2       \\
relativistic scalar $(\alpha^{\rm S})$ $^{\rm\ a}$   &  97\,101 &6  \\
Relativistic tensor $(-\frac12\alpha^{\rm T})$ &   1\,744& \\
\hline
Total non-QED        & 725\,743\,960&6     \\
\hline
\multicolumn{3}{c}{\textbf{QED terms}} \\
QED $\alpha^3$       &       -7\,297& 2       \\
QED $\alpha^4$       &          -127& 6       \\
\hline
Total QED            &         -7\,424& 6     \\
\hline
Retardation          &          -477 &        \\
Nuclear size $^{\rm\ b}$         &             5&          \\
Magnetic polarizability&         188&         \\
\hline
Grand total          & 725\,736\,252& 9       \\
Experiment           & 725\,736\,700& 260 \\
\hline
Difference           &          -448& 260\\
\end{tabular}\\
\end{ruledtabular}
$^{\rm a\,}$The value of the Rydberg used is $3.289\,841\,960\,2508(64)\times10^9$ MHz.  \\
$^{\rm b\,}$This value is converted from data 2.75~fm in Ref.~\cite{PhysRevA.99.040502}.
\end{table}
In addition, the leading QED corrections are included according to the effective operators \cite{Yerokhin2010}
\begin{equation}
\hat{X}_{\rm QED}^{(3)} = \alpha^3\left\{\frac{4Z}{3}\left(\frac{19}{30} - \ln\alpha^2 -
\ln k_0\right)[\delta^3({\bf r}_1) + \delta^3({\bf r}_2)]
 - \frac{7}{6\pi}\left(\frac{1}{r_{12}^3}\right)_{\rm P.V.}\right\},
\end{equation}
and
\begin{equation}
\hat{X}_{\rm QED}^{(4)} =\alpha^4\pi\left[\left(\frac{427}{96} - 2\ln2\right)Z^2 + \left(-\frac{9\zeta(3)}{4\pi^2} - \frac{2179}{648\pi^2} + \frac{3\ln2}{2} - \frac{10}{27}\right)Z\right]
[\delta^3({\bf r}_1) + \delta^3({\bf r}_2)], \label{alpha4}
\end{equation}
where $\zeta(z)$ is the Reimann zeta-function, and $(1/r_{12}^3)_{\rm P.V.}$ denotes the principal value of the divergent integral, as defined by
\begin{equation}
\left(\frac{1}{r_{12}^3}\right)_{\rm P.V.} = \lim_{\epsilon\rightarrow 0}
r_{12}^{-3}(\epsilon)+ 4 \pi (\gamma +\ln \epsilon) \delta({\rm\bf r}_{12}),
\end{equation}
where $\epsilon$ is the radius of a small sphere about $r_{12} = 0$ that is omitted from the range of integration, and $\gamma$ is Euler's constant (the final result is independent of $\gamma$).  Here, $\hat{X}_{\rm QED}^{(3)}$ contains the Araki-Sucher terms \cite{Araki,Sucher}, and $\hat{X}_{\rm QED}^{(4)}$ contains the radiative QED terms of order $\alpha^4$ Ry.  We take the remaining nonradiative contribution of about 5\% of the radiative terms for the $2^3S_1$ state \cite{Piszczatowski2015,Pachucki2006} (i.e. about 6~MHz) to be the dominant source of uncertainty.

The term $\delta\bar{\alpha}_{\rm d}^{\partial_{\cal E}^2\ln k_0}(\omega)$ in Eq.~(\ref{eq:baralpha}) represents the second-order electric field perturbation to the Bethe logarithm $\ln k_0$.  This term has recently been calculated by Puchalski \textit{et al.}~ \cite{PhysRevA.101.022505} for the ground state of helium, with the result $\delta\bar{\alpha}_{\rm d}^{\partial_{\cal E}^2\ln k_0}(0) = 0.048\,557\,2(14)$ $a_0^3$.  However, as pointed out by Drake and Yan \cite{DrakeBL}, the Bethe logarithm (expressed in the $Z$-scaled form $\ln(k_0/Z^2)$) is determined almost entirely by the inner $1s$ electron, and so is nearly independent of the atomic state, or even the ionization state, at the $\pm1$\% level of accuracy.  The same is also true for the finite nuclear mass corrections to the Bethe logarithm. We can therefore safely assume that $\delta\bar{\alpha}_{\rm d}^{\partial^2_{\cal E}\ln k_0}(\omega) \simeq 0.049(1)$ $a_0^3$, independent of $\omega$ for small $\omega$.  The corresponding correction to the tune-out frequency is then
\begin{equation}
\delta\omega_{\rm TO} = -\frac{8\alpha^3}{3}\delta\bar{\alpha}_{\rm d}^{\partial^2_{\cal E}\ln k_0}
\langle \delta^3({\bf r}_1) + \delta^3({\bf r}_2)\rangle\left(\frac{1}{\partial\bar{\alpha}_{\rm d}(\omega)/\partial\omega|_{\omega_{\rm TO}}}\right).
\end{equation}
Using the values $\langle2^3S_1|\delta^3({\bf r}_1) + \delta^3({\bf r}_2)|2^3S_1\rangle = 8.29604/\pi$ $a_0^{-3}$ and $\partial\bar{\alpha}_{\rm d}(\omega)/\partial\omega|_{\omega_{\rm TO}}$ = 7134 $\hbar a_0^4/e^2$, the numerical value is $\delta\omega_{\rm TO} \simeq 0.124(3)$ MHz, which is negligible at current levels of accuracy.  The dominant source of uncertainty is therefore the nonradiative QED corrections of order $\alpha^4$ Ry, which we take to be $\pm6$ MHz, as explained in the previous paragraph.

The magnetic susceptibility is defined by \cite{Bruch2003,Pachucki_Yerokhin}
\begin{eqnarray}
\chi = -\frac{\alpha^2a_0}{4}\langle \sum_{i=1}^2({\bf r}_i\times\hat{\bf B})^2\rangle= -\frac{\alpha^2a_0}{6}\langle \sum_{i=1}^2r_i^2\rangle\,, \quad \mbox{for $S$-states}
\end{eqnarray}
and $\hat{\bf B}$ denotes a unit vector.  For the $2\,^3S$ state, $\langle \sum_{i=1}^2r_i^2\rangle = 22.928\,644$ $a_0^2$ for the case of infinite nuclear mass.  This produces a frequency-independent offset when calculating the tune-out frequency, resulting in a correction of 188 MHz as shown in Table~\ref{tab:theory}.

A major new contribution of the present work is to include the retardation corrections recently derived by Pachucki and Puchalski \cite{PhysRevA.99.041803,Drake2019} and evaluated for the ground state of helium.  These terms represent a reformulation of the problem as a zero in the coherent Rayleigh scattering amplitude for an atom in free space, instead of a zero in the frequency-dependent polarizability for an atom in an optical lattice \cite{PhysRevA.99.041803,Drake2019}.  Here we extend the calculations to the $2^3S_1$ state of helium.  The generalized polarizabilities are defined by (in units of $a_0^5$, except for \(\alpha_0\) which is in units of $a_0^3$)
\begin{eqnarray}
\alpha_0(\omega) &=& \frac{e^2}{3}\sum_{a,b}\langle r_a^k{\cal R}(\omega)r_b^k\rangle\\
\alpha_1(\omega) &=& \frac{e^2}{3}\sum_{a,b}\langle (r_a^kr_a^l)^{(2)}{\cal R}(\omega)(r_b^kr_b^l)^{(2)}\rangle\\
\alpha_2(\omega) &=& \frac{e^2}{3}\sum_{a,b}\langle r_a^k{\cal R}(\omega)r_b^kr_b^2\rangle\\
\alpha_3(\omega) &=& \frac{2\hbar e^2}{3m}\sum_{a,b}\langle r_a^k{\cal R}(\omega){\cal R}(\omega)i({\rm L_b}\times{\rm r}_b - {\rm r}_b\times{\rm L}_b)^k\rangle\\
\alpha_4(\omega) &=& \frac{e^2}{3}\sum_{a,b}\langle r_a^2{\cal R}(\omega)r_b^2\rangle,
\end{eqnarray}
where the $a$ and $b$ sums are over the electronic coordinates 1 and 2, and a sum over the repeated cartesian vector components $k$ and $l$ is assumed.  Also, $(r_a^kr_a^l)^{(2)} \equiv r^kr^l -  \delta_{k,l}r^2/3$ is the quadrupole moment operator.  The above definitions differ by a factor of 2/3 from those in Ref.\ \cite{PhysRevA.99.041803} so that here $\alpha_0(\omega) \equiv \alpha_{\rm d}(\omega)$, and $\alpha_1(\omega)$ corresponds to the standard definition of the quadrupole polarizability \cite{Ho2020}.  For each term, $\bar{\alpha}_i(\omega) = \frac12[\alpha_i(\omega) + \alpha_i(-\omega)]$. The retardation corrections to the polarizability $\alpha_0$ and diamagnetic coupling $\chi$ are then
\begin{eqnarray}
\bar{\alpha}_{\rm ret}(\omega) &=& \frac{3k^2}{2}\left(\frac{\bar{\alpha}_1(\omega)}{15} - \frac{2\bar{\alpha}_2(\omega)}{15} + \frac{\bar{\alpha}_4(\omega)}{18}\right)\\
\bar{\chi}_{\rm ret}(\omega) &=& \frac{3k^2}{2}\left(-\frac{\bar{\alpha}_1(\omega)}{60} + \frac{4\bar{\alpha}_2(\omega)}{45} + \frac{\bar{\alpha}_3(\omega)}{9}
- \frac{\bar{\alpha}_4(\omega)}{18}\right),
\end{eqnarray}
where $k = \omega/c$.  The total retardation correction to $\alpha_{\rm d}(\omega)$ is
\begin{equation}
\bar{\alpha}_{\rm ret}(\omega) + \bar{\chi}_{\rm ret}(\omega)
= \frac{3k^2}{2}\left(\frac{\bar{\alpha}_1(\omega)}{20} - \frac{2\bar{\alpha}_2(\omega)}{45} + \frac{\bar{\alpha}_3(\omega)}{9}\right).
\end{equation}
As shown in Table~\ref{tab:theory}, the total retardation correction is $-477$ MHz.  

In addition to the above, there are two possible QED corrections to the dipole transition operator.  The first, discussed by Pachucki \cite{Pachucki2004} is given by
\begin{equation}
\delta{\bf r}_{\rm QED} = \frac{i\hbar \omega\kappa}{4mc^2}\sum_{i=1}^2 {\bf r}_i\times\sigma_i
\end{equation}
where $\kappa = \alpha/(2\pi)$ is the anomalous magnetic moment.  However, if the fine-structure splittings of the intermediate $n\;^3P_J$ states are neglected in the sum over states, then this term sums to zero, as can be seen from the orthogonality property of the 6-$j$ symbols \cite{Edmonds}.  The residual contribution is suppressed by a factor of $\alpha^2$ and can be neglected.  The other relativistic and retardation terms in Eq.\ (51) of Ref.\ \cite{Pachucki2004} are already included in the $\alpha_2$ and $\alpha_3$ terms.

For triplet states, there is a second QED vertex correction to the dipole transition operator from the anomalous magnetic moment of the electron.  It can be obtained from the effective Dirac operator \cite{Akhiezer} ${\cal D} = \mu_0\alpha/(2\pi)(\beta\mbox{\boldmath{$\Sigma\cdot{\bf B}$}} - i\beta\mbox{\boldmath{$\alpha\cdot {\bf E}$}})$, where $\mu_0$ is the Bohr magneton $e\hbar/2mc$.  Cross terms of the form $\langle i |\mbox{\boldmath{$\alpha\cdot{\bf A}$}} | n\rangle\langle n|{\cal D} | i\rangle$ vanish due to transversality for the case of coherent Rayleigh scattering.  Matrix element products of the form $\mu_0^2\langle i|\beta\mbox{\boldmath{$\Sigma\cdot{\bf B}$}}|n\rangle\langle n|\beta \mbox{\boldmath{$\Sigma\cdot{\bf B}$}}|i\rangle$ of relative order $\alpha^4$ have a zero at the same tune-out frequency as the main term, and so they do not change the tune-out frequency.

For purposes of practical calculations, the resolvent operator is expanded in terms of pseudostates so that the polarizability assumes its familiar form
\begin{equation}
\bar{\alpha}_{\rm d}(\omega) = e^2\sum_n |\langle i|\edotr |n\rangle|^2\left(\frac{1}{E_n-E_i+\hbar\omega} +
\frac{1}{E_n-E_i-\hbar\omega}\right),
\end{equation}
and similarly for all the other correction terms.  The basis sets in correlated Hylleraas coordinates are all of the double basis set form \cite{Drake88,Drake92}
\begin{equation}
\Psi({\bf r}_1,{\bf r}_2) = \sum_{t=1}^2\sum_{i,j,k}^{i+j+k\le\Omega}
a_{i,j,k}^{(t)} r_1^ir_2^jr_{12}^k\exp(-\alpha^{(t)}r_1 - \beta^{(t)}
r_2)\, r_1^{l_1}r_2^{l_2}\, {\cal Y}_{l_1l_2L}^M(\hat{\bf r}_1,\hat{\bf r}_2) - {\rm exchange},
\end{equation}
where $t$ labels two independent sets of nonlinear parameters $\alpha^{(1)}, \beta^{(1)}$ and $\alpha^{(2)}, \beta^{(2)}$ that set two distinct distance scales, and are individually optimized for the state in question, together with the accompanying set of pseudostates.  ${\cal Y}_{l_1l_2L}^M(\hat{\bf r}_1,\hat{\bf r}_2)$ denotes a vector coupled product of spherical harmonics.  The numerical uncertainty in the final results was determined by systematically increasing the parameter $\Omega$ up to $\Omega = 16$ and studying the convergence.  All calculations were done in standard quadruple precision arithmetic (about 32 decimal digits).

\subsection{Theoretical Results}
The results are as indicated in Table~\ref{tab:theory}. The various entries from nonrelativistic, relativistic, and QED contributions are not strictly additive because changing one effect, such as the relativistic correction, changes the tune-out frequency at which the other effects are evaluated.  Thus the first entry is the nonrelativistic tune-out frequency with finite nuclear mass effects included.  The next entry is the scalar part of the relativistic correction arising from the $p^4$, $H_2$ and $H_4$ terms in the Breit interaction, and iterated to convergence.  The next entry from the $H_5$ spin-spin interaction (1743.8~MHz) is solely and exclusively responsible for the tensor part of the tune-out frequency (excluding the Schwinger radiative correction term $\alpha/\pi$).  These terms determine the nonrelativistic plus relativistic part of the tune-out frequency, iterated to convergence.  The remaining terms are small enough that they can be added linearly.  The leading QED correction of order $\alpha^3$ Ry ($-7296.6(5)$ MHz) includes the anomalous magnetic moment correction to $H_5$ (8 MHz) and the very small estimate of $-0.124(3)$ MHz for the $\delta\bar{\alpha}_{\rm d}^{\partial_{\cal E}^2\ln k_0}$ term.  The terms of order $\alpha^4$ Ry include only the radiative corrections in Eq.~(\ref{alpha4}).  The dominant source of uncertainty is thus the remaining nonradiative terms not included in the present calculation, but evaluated previously for the $2^3S_1$ state energy \cite{Pachucki2006}.  The remaining terms are the retardation correction of $-477$ MHz, evaluated here for the first time for the helium $2^3S_1$ state, and a finite nuclear size correction.

The final theoretical result of 725\,736\,252(9) MHz by the nr-QED method differs from experiment by $-448\pm260$ MHz.  A more detailed account of the theory and calculations will be published separately.

\section{Comparison With Previous Oscillator Strength Ratio Measurements}
\label{sec:osc_ratio}

We claim that this measurement of the $\TO$ tune-out wavelength represents the most precise measurement of transition rate ratios made to date in an atomic system. To find the uncertainty in the ratio of oscillator strengths we start by generalizing the treatment given in Ref.~\cite{PhysRevA.88.052515} by using a model of the polarizability of a three level system.
\begin{equation}
\alpha(f) = \frac{\mathfrak{F}_1}{E_1^2-h^2 f^2}+\frac{\mathfrak{F}_2}{E_2^2-h^2 f^2}
\end{equation}
where $\mathfrak{F}_1$, $\mathfrak{F}_2$ correspond to the oscillator strengths, and $E_1$, $E_2$ the excitation energy of the respective transitions, $f$ is the photon frequency, and \(h\) is Planck's constant. 

If we introduce  the ratio of the oscillator strengths $X=\mathfrak{F}_2^2/\mathfrak{F}_1^2$ and note that by definition \(\alpha(f_{TO})=0\), substituting in the above expression and solving for this ratio we find
\begin{equation}
    X =\frac{ (E_2^2 -h^2 f_{TO}^2 )^2}{(E_1^2-h^2 f_{TO}^2)^2}.
\end{equation}
We can hence find the sensitivity of the value of \(X\) to changes in the tune-out frequency.
\begin{equation}
    \frac{\delta X}{X} = \frac{\partial X} {\partial f_{TO}} \frac{1}{X} \cdot \delta f_{TO} =
    \frac{2  h^2 f_{TO} (E_1^2-E_2^2)}
    {(E_1^2-h^2 f_{TO}^2) (-E_2^2 + h^2 f_{TO}^2 )} \cdot \delta  f_{TO} = \frac{-2 f_{TO}^2 (f_1^2-f_2^2)}{(f_1^2-f_{TO}^2)(f_2^2-f_{TO}^2)} \frac{\delta  f_{TO}}{f_{TO}}
\end{equation}
where \(E_1=h f_1\) and \(E_2=h f_2\).
To evaluate other work in the literature we use the frequency of the dominant transitions and the measured tune-out value to then derive the estimated sensitivity to the ratio of transition strengths. This method is approximate and neglects the contribution from the DC polarizability, however this is a small effect and not needed for such coarse comparison of sensitivity. Given the dominant transition manifolds at 276.7465~THz (\(\MetastableState \to \UpperStateManifold \)) and 770.7298~THz (\(\MetastableState \to \LowerStateManifold\)), as well as our value for the tune-out frequency of \(f_{TO}=725.7367\)~THz, we reach a fractional uncertainty in the oscillator strength ratio of 6~ppm, an improvement on the previous record of 15~ppm set by Ref.~\cite{PhysRevA.92.052501}. We note that the fractional uncertainty in the ratio of oscillator strengths is identical to the fractional uncertainty in the ratio of transition matrix elements which are calculated in Ref.~\cite{PhysRevA.92.052501}.

\end{document}